\documentclass{scrartcl}
\pdfoutput=1

\usepackage{hyperref}

\hypersetup{
	pdfauthor={Boris Kramer, Alexandre Noll Marques, Benjamin Peherstorfer, Umberto Villa, and Karen Willcox},
	pdftitle={Multifidelity probability estimation via fusion of estimators},
	pdfkeywords={Multifidelity modeling, uncertainty quantification, information fusion, importance sampling, reduced-order modeling, failure probability estimation, PDE, turbulent jet},
	pdfproducer={Latex with hyperref}}

\usepackage{amsmath}
\usepackage{amsfonts}
\usepackage{amssymb}
\usepackage{amsthm}
\usepackage[usenames,dvipsnames]{xcolor}
\usepackage{graphicx}
\usepackage{url}
\usepackage{subfigure}
\usepackage{geometry}
\usepackage{algorithmic,algorithm}  

\newcommand{\Sig}{\boldsymbol{\Sigma}}
\newcommand{\limitg}{g}
\newcommand{\failset}{\mathcal{G}}
\newcommand{\param}{\mathbf{z}}
\newcommand{\paramdom}{\mathcal{D}}
\newcommand{\RV}{Z}
\newcommand{\state}{s}
\newcommand{\stateSpace}{\tilde{\Omega}}
\newcommand{\NL}{\mathcal{F}}
\renewcommand{\Pi}{P_i}
\newcommand{\Pj}{P_j}
\newcommand{\Palpha}{P_{\boldsymbol{\alpha}}}
\newcommand{\Phatalpha}{\hat{P}_{\boldsymbol{\alpha}}}

\newtheorem{problem}{Problem}
\newtheorem{corollary}{Corollary}
\newtheorem{proposition}{Proposition}

\newenvironment{keywords}%
{\begin{trivlist}\item[]{\bfseries\sffamily Keywords: }\ }
	{\end{trivlist}}

\title{Multifidelity probability estimation \\ via fusion of estimators}

\author{
	Boris Kramer \thanks{Department of Aeronautics \& Astronautics, Massachusetts Institute of Technology, Cambridge, MA {\url{bokramer@mit.edu}}, \url{noll@mit.edu}.}
	\and Alexandre Noll Marques\footnotemark[1]
	\and Benjamin Peherstorfer\thanks{Courant Institute of Mathematical Sciences, New York University, NY (\url{pehersto@cims.nyu.edu}).}
	\and Umberto Villa\thanks{Department of Electrical and Systems Engineering, Washington University in St. Louis, MO (\url{uvilla@wustl.edu}).}
	\and Karen Willcox\thanks{Oden Institute for Computational Engineering \& Sciences, The University of Texas at Austin, TX (\url{kwillcox@ices.utexas.edu})}
}
\date{April 22, 2019}

\begin{document}

\maketitle

\begin{abstract}
\noindent
\small{
This paper develops a multifidelity method that enables estimation of failure probabilities for expensive-to-evaluate models via  information fusion and importance sampling.
The presented general fusion method combines multiple probability estimators with the goal of variance reduction. 
We use low-fidelity models to derive biasing densities for importance sampling and then fuse the importance sampling estimators such that the fused multifidelity estimator is unbiased and has mean-squared error lower than or equal to that of any of the importance sampling estimators alone.
By fusing all available estimators, the method circumvents the challenging problem of selecting the best biasing density and using only that density for sampling. 
A rigorous analysis shows that the fused estimator is optimal in the sense that it has minimal variance amongst all possible combinations of the estimators. 
The asymptotic behavior of the proposed method is demonstrated on a convection-diffusion-reaction partial differential equation model for which $10^5$ samples can be afforded. 
To illustrate the proposed method at scale, we consider a model of a free plane jet and quantify how uncertainties at the flow inlet propagate to a quantity of interest related to turbulent mixing. 
Compared to an importance sampling estimator that uses the high-fidelity model alone, our multifidelity estimator reduces the required CPU time by 65\% while achieving a similar coefficient of variation.}
\end{abstract}

\section{Introduction}
This paper considers multifidelity estimation of failure probabilities for large-scale applications with expensive-to-evaluate models. 
	Failure probabilities are required in, e.g.,  reliable engineering design and risk analysis. Yet failure probability estimation with expensive-to-evaluate nonlinear models is computationally challenging due to the large number of Monte Carlo samples needed for low-variance estimates.
	
	Efficient failure probability estimation methods aim to reduce the number of samples at which the expensive model is evaluated, e.g., by exploiting variance-reducing sampling strategies, multifidelity/multilevel estimation methods, or sequential sampling approaches.
	Variance reduction can be obtained through importance sampling \cite{owen2013monte}, which allows for order-of-magnitude reductions in the number of samples needed to reliably estimate a small probability. However, importance sampling relies on having a good biasing distribution which in turn requires insight into the system. Surrogate models can provide such insight at much lower computational cost. Multifidelity approaches (see \cite{PWG17MultiSurvey} for a review) that use surrogates for failure probability estimation via sampling have seen great interest recently \cite{dongbin2011,chen_accurate_2013, dongbin2014, Elferson16_MLMCFailure,Ullmann15_MultilevelSubset,Peherstorfer16MFIS,Fagerlund16_MLMCporousflows}, but require that the user selects a good importance sampling density. Multifidelity methods that avoid the selection of a single biasing density and instead use a suite of surrogate models to generate importance sampling densities were proposed in \cite{PKW16MixedMFIS,PKW17MFCE,OwenIS}. Nevertheless, this framework requires all knowledge about the small probability event to be available in the form of biasing densities, and is therefore only applicable to importance sampling estimators. 
	Multilevel Monte Carlo~\cite{giles2008multilevel,aslett2017multilevel} methods use a hierarchy of approximations to the high-fidelity model in the sampling scheme. However, those model hierarchies have to satisfy certain error decay criteria, an assumption we do not make here.  Subset simulation~\cite{AuBeck_2001SS,papaioannou2015mcmcSS} and line search~\cite{schueller2004critical,angelis2015advanced} can be used directly on the high-fidelity models, and therefore are of a black-box nature. 
	
	In this work, in addition to the computationally expensive model, we also have information about the system in form of surrogate models, analytical models, expert elicitation, and reduced models.  In other settings where such a variety of information is available, information fusion has been used to combine multi-source probabilistic information into a single estimator, see  \cite{clemenWinkler99combiningExperts, Ohagan06ElicitingExpProb, marin10weightingVariance}. Moreover, combining information from multiple models and sources via a weighted multifidelity method can lead to efficient data assimilation strategies \cite{NARAYAN2012}.

Here, we propose a new approach to enable small probability estimation for large-scale, computationally expensive models that draws from prior work in information fusion, importance sampling, and multifidelity modeling. 
We use information fusion in combination with multifidelity importance-sampling-based failure probability estimators, where in addition to the variance reduction from importance sampling, we obtain further variance reduction through information fusion.	
The proposed multifidelity framework uses the available surrogates to compute multiple unbiased failure probability estimators. We then combine them \textit{optimally} into a new unbiased estimator that has minimal variance amongst all possible linear combinations of those estimators. 
The method therefore avoids the selection of the lowest variance biasing density to be used for sampling. Selecting the density that leads to the lowest variance in the failure probability estimator would require additional information, and not even error estimates on the surrogate model would suffice. Thus, we circumvent this step and optimally use all information available to us in form of probability estimators. 

This paper is structured as follows: In Section~\ref{sec:lowprob} we illustrate the challenges in small failure probability computation and cover the necessary background material for multifidelity importance sampling used herein. Section~\ref{sec:CompMethods} details our proposed approach of information fusion, importance sampling and multifidelity modeling. We then present in Section~\ref{sec:CDR}  a moderately expensive convection-diffusion-reaction test case, where we illustrate the asymptotic behavior of our approach. Section~\ref{sec:resultsjet} discusses a turbulent jet model and demonstrates the computational efficiency of our proposed methods for this computationally expensive model. We close with conclusions in Section~\ref{sec:conclusions}.

\section{Small probability events and importance sampling estimators} \label{sec:lowprob}
We are interested in computing events with small probabilities, e.g., failure events, where the system fails to meet critical constraints. Section~\ref{sec:SmallProb} describes small probability events, Section~\ref{sec:IS} introduces importance sampling and Section~\ref{sec:MFIS} briefly summarizes multifidelity importance sampling.

\subsection{Small probability events} \label{sec:SmallProb}
Let $\Omega$ be a sample space which, together with a sigma algebra and probability measure, defines a probability space. 
Define a $d$-dimensional random variable $\RV: \Omega \mapsto \paramdom \subseteq \mathbb{R}^{d}$ with probability density $p$, and let $\param$ be a realization of $\RV$. 
Let $f: \paramdom \subseteq \mathbb{R}^d \mapsto \mathbb{R}^{d'}$ be an expensive-to-evaluate model of high fidelity with corresponding $d'$-dimensional quantity of interest $f(\param)\in \mathbb{R}^{d'}$. Let $g: \mathbb{R}^{d'}\mapsto \mathbb{R}$ denote a limit state function that defines failure of the system. 
If $\limitg(f(\param))<0$, then $\param \in \paramdom$ is a configuration where the system fails. 
This defines a failure set 
\begin{equation*}
	\failset:= \{ \param \in \paramdom \  | \ \limitg(f(\param))<0 \}.
\end{equation*}
Define the indicator function $I_\failset : \paramdom \mapsto \{ 0,1 \}$ via 
\begin{equation*}
	I_\failset(\param) = \begin{cases}
		1\,,\qquad \param \in \failset\, ,\\
		0\,,\qquad \text{otherwise}\,.
	\end{cases}
\end{equation*}
The standard Monte Carlo estimator of the failure probability 
$$
P = \mathbb{E}_p [I_\failset[\RV]] = \int_{\paramdom} I_\failset(\param) p(\param) d\param
$$ 
uses $n$ realizations $\param_1, \ldots, \param_n$ of the random variable $\RV$ and estimates
\begin{equation}
	P_n = \frac{1}{n}\sum_{i=1}^n I_\failset (\param_i). \label{eq:plainMCest}
\end{equation}
In the special case of small probabilities, standard Monte Carlo may be unfeasible due to the large number of samples needed to obtain good estimators.
Since failure probabilities are generally small, most realizations $\param_i$ will be outside the failure domain $\failset$, and conversely, only a small fraction of the $n$ samples lies in the failure region.  
The coefficient of variation (also called relative root-mean-squared error) of the estimator $P_n$ is given by 
\begin{equation}
	e^{\text{CV}} (P_{n}) = \sqrt{\frac{\mathbb{V}[P_n]}{(\mathbb{E}[P_n])^2} } =  \sqrt{\frac{P(1-P)}{n P^2} } = \sqrt{\frac{1-P}{n P} } . \label{eq:CVwithP}
\end{equation}
Thus, to obtain estimators with a small coefficient of variation, a large number of samples is necessary. For instance, if the small probability is $P=10^{-4}$ and  if we want $e^{\text{CV}}=10^{-1}$ we would need $n=\mathcal{O}(10^{6})$ samples via standard Monte Carlo approaches. This challenge is amplified by the presence of an expensive-to-evaluate model, such as the model of a free plane jet in Section~\ref{sec:resultsjet}.

\subsection{Importance sampling} \label{sec:IS}
Importance sampling achieves variance reduction by using realizations of a random variable $\RV': \Omega \mapsto \paramdom$ with probability density $q$. This random variable $\RV'$ is chosen such that its probability density function $q$ has higher mass (compared to the nominal density $p$) in the region of the event of interest.  For a general introduction to importance sampling, see \cite[Sec.9]{owen2013monte}. 
Define the support $\text{supp}(p)=\{ \param \in \paramdom \ | \ p(\param)>0 \}$, and let $\text{supp}(p)\subseteq \text{supp}(q)$. Then 
\begin{equation}
	P = \int_{\paramdom} I_\failset(\param) p(\param) d\param  = \int_{\paramdom} I_\failset(\param) \frac{p(\param)}{q(\param)} q(\param) d \param \label{eq:Pint}
\end{equation}
is well defined, where $p(\param) /  q(\param)$ is the \textit{likelihood ratio}---in the context of importance sampling also called \textit{importance weight}.
The importance-sampling estimate of the failure probability $P$ then draws $n$ realizations $\param'_1, \ldots, \param'_n$ of the random variable $\RV'$ with density $q$ and evaluates 
\begin{equation}
	P_n^{\text{IS}} = \frac{1}{n} \sum_{i=1}^n I_\failset (\param'_i) \frac{p(\param'_i)}{q(\param'_i)}. \label{eq:IS}
\end{equation}
The variance of the importance sampling estimator is 
\begin{equation}
	\mathbb{V}[P_n^{\text{IS}}] = \frac{\sigma_q^2}{n}, \label{eq:VarIS}
\end{equation}
where 
\begin{equation}
	\sigma_q^2 = \int_\mathcal{D} \left ( \frac{I_\failset (\param') p(\param') }{q(\param')} - P \right )^2 q(\param')d\param' . \label{eq:ISasympVar}
\end{equation}
If  $\text{supp}(p)\subseteq \text{supp}(q)$, and by using \eqref{eq:Pint}, one can show that the importance sampling estimator $P_n^{\text{IS}}$ is an unbiased estimator of the failure probability, i.e., 
\begin{equation*}
	\mathbb{E}_q [P_n^{\text{IS}}] = \mathbb{E}_p [I_\failset(\RV)] = P.
\end{equation*}
The importance sampling estimator $P_n^{\text{IS}}$ has mean $P$ and variance $\sigma_q^2/n$, and by the central limit theorem converges in distribution to the normal random variable $\mathcal{N}(P,\sigma_q^2/n)$.
Constructing a good biasing density that leads to small $\sigma_q^2$ is challenging~\cite{owen2013monte}. We next introduce low-fidelity surrogate models, which are then used to construct biasing densities.

\subsection{Multifidelity Importance Sampling (MFIS)} \label{sec:MFIS}
Recall that by $f:\paramdom \mapsto \mathbb{R}^{d'}$ we denote an expensive-to-evaluate model of high fidelity with corresponding quantity of interest $f(\param)\in \mathbb{R}^{d'}$. Let $k$ surrogates 
\begin{equation*}
	f^{(i)}: \paramdom \mapsto \mathbb{R}^{d'}, \ i=1, \ldots, k
\end{equation*}
of lower fidelities be available, which are cheaper to evaluate than the high-fidelity model $f(\cdot)$. We do not assume any information about the accuracy of the $f^{(i)}(\cdot)$ with respect to the high-fidelity model $f(\cdot)$. Sections~\ref{sec:CDRdiscrete}  and \ref{sub:jet_surrogate} detail the specific surrogate models used for the respective applications. 

We use the MFIS method (see~\cite{Peherstorfer16MFIS} for details) to obtain $k$ estimators of the failure probability. First, MFIS evaluates the surrogate models $f^{(i)}$ at $m_i$ samples to obtain a surrogate-model specific failure set $\failset^{(i)}$. Second, MFIS computes a biasing density $q_i$ by fitting a distribution in form of a Gaussian mixture model to the parameters in the failure set. If no failed samples are found by the surrogate model, i.e., if $\failset^{(i)}=\emptyset$, then we set the biasing density to be the nominal density. This  leads to $k$ biasing densities $q_1, \ldots, q_k$ from which we get importance sampling estimators
\begin{equation} \label{eq:PniIS}
	P_{n_i}^{\text{IS}} = 
	\frac{1}{n_i} \sum_{j=1}^{n_i} I_\failset (\param_{i,j}) \frac{p(\param_{i,j})}{q_i(\param_{i,j})}, \qquad \param_{i,j}\sim q_i, \ j=1,\ldots, n_i, 
\end{equation}
for $i=1,\ldots, k $. The variance of the importance sampling estimator is given by \eqref{eq:VarIS} with $n=n_i$ and $\sigma_q=\sigma_{q_i}$, with $\sigma_{q_i}^2$ being the asymptotic variance from \eqref{eq:ISasympVar} with $q=q_i$.

\section{Fusion of multifidelity estimators} \label{sec:CompMethods}
In many practical situations, a range of probability estimators are available, for instance in form of MFIS estimators derived from different biasing densities, in form of analytical models, or estimators derived from expert elicitation~\cite{Ohagan06ElicitingExpProb}. 
If one a priori knew which was the lowest variance estimator then a good strategy would be to sample only from that estimator. However, knowing \textit{a priori} which estimator has the lowest variance is a formidable task, and one has to draw samples to assess which estimator has the lowest variance.
In this section, we present our new approach that combines all available estimators in an optimal fashion by solving the following problem.

\begin{problem} \label{probdef}
	Given $k$ unbiased estimators, $P_1, \ldots, P_k$ with expected value $P$, i.e. $\mathbb{E}[\Pi] = P, \ i=1, \ldots, k$, find an estimator with expected value $P$ of the form 
	\begin{equation}
		\Palpha = \sum_{i=1}^k \alpha_i \Pi,  \label{eq:Palpha} 
	\end{equation}
	such that it attains minimal variance amongst all estimators of the form \eqref{eq:Palpha}. That is, find the optimal weights $\alpha_i \in \mathbb{R}, \ i =1,\ldots,k$ such that
	\begin{equation}
		\min_{\boldsymbol{\alpha}} \ \mathbb{V}[P_{\boldsymbol{\alpha}}] \quad \textup{s.t.} \quad \mathbb{E}[P_{\boldsymbol{\alpha}}] = P. \label{eq:minprob}
	\end{equation}
\end{problem}
\noindent
The fused estimator approach allows to still use information coming from the other (high-variance) estimators, whose samples would have otherwise gone to waste. Moreover, with the proposed method we can estimate small-probabilities for expensive-to-evaluate models by exploiting a variety of surrogates. 
We derive expressions for the mean and variance of the fused estimator in Section~\ref{sec:meanVar}. In Section~\ref{sec:OptWeights}, we derive the optimal weights for the fused estimator. Section \ref{sec:Uncorr} then discusses the special case of uncorrelated estimators. Our proposed algorithm is discussed in Section~\ref{sec:fusedIS}, followed by a brief Section~\ref{sec:errComp} that discusses measures of convergence of the estimators.

\subsection{Mean and variance of fused estimator}\label{sec:meanVar}
We start with the observation that if the weights $\alpha_i$ of the fused estimator $\Palpha$  sum to one, then the fused estimator  is unbiased:
\begin{equation*}
	\sum_{i=1}^k \alpha_i = 1 \quad \Leftrightarrow \quad \mathbb{E}[\Palpha] = \sum_{i=1}^k \alpha_i \mathbb{E}[\Pi] = P \sum_{i=1}^k \alpha_i = P.  
\end{equation*}
Let the estimators $P_i$ have corresponding variances $0< \sigma_{i}^2<\infty, \ i=1,\ldots k$. To compute the variance of the fused estimator $P_{\boldsymbol{\alpha}}$ we have to consider covariances between the individual estimators. 
Define the Pearson product-moment correlation coefficient as 
\begin{equation}
	\rho_{i,j} = \frac{\mathbb{C}\text{ov} (\Pi,\Pj)}{\sigma_i \sigma_j}, \label{eq:CorrCoef}
\end{equation}
where $\mathbb{C}\text{ov}(\Pi,\Pj)=\mathbb{E}[(\Pi- \mathbb{E}[\Pi])(\Pj- \mathbb{E}[\Pj])] = \mathbb{E}[\Pi \Pj] - P^2$.
We also define the symmetric, positive semi-definite \textit{covariance matrix} $\Sig_{ij} = \mathbb{C}\text{ov}(\Pi,  \Pj)$ as: 
\begin{equation}
	\Sig = \begin{bmatrix}
		\sigma_{1}^2 & \sigma_{1}\sigma_{2} \rho_{1,2} &\ldots  & \ldots & \sigma_{1}\sigma_{k} \rho_{1,k} \\
		\sigma_{2}\sigma_{1} \rho_{2,1} & \sigma_{2}^2 & \sigma_{2}\sigma_{3} \rho_{2,3} & \ldots & \sigma_{2}\sigma_{k} \rho_{2,k} \\
		\vdots & \ddots &  & & \\
		\vdots & & & \sigma_{{k-1}}^2 & \sigma_{{k-1}}\sigma_{k} \rho_{k-1,k} \\
		\sigma_{k}\sigma_{1} \rho_{k,1} & \sigma_{k}\sigma_{2} \rho_{k,2} & \ldots & \sigma_{{k}}\sigma_{{k-1}} \rho_{k,k-1} & \sigma_{k}^2 \\
	\end{bmatrix}. \label{eq:Mmatrix}
\end{equation}
It is worth noticing that if the estimators $P_1,\ldots, P_k$ are independent, then $\Sig$ is diagonal. The variance of the fused estimator from \eqref{eq:Palpha} is 
\begin{align*}
	\mathbb{V}[\Palpha]  = \mathbb{V} \left [ \sum_{i=1}^k \alpha_i \Pi \right ] 
	& = \sum_{i=1}^k \alpha_i^2 \mathbb{V}[\Pi] + 2 \mathbb{C}\text{ov} \left ( \sum_{i=1}^k \alpha_i \Pi, \, \sum_{j=1}^k \alpha_j \Pj \right ) \\
	& = \sum_{i=1}^k \alpha_i^2 \sigma_i^2 + 2 \sum_{i=1}^k \sum_{j>i}^{k} \alpha_i \alpha_j \sigma_i \sigma_j \rho_{i,j}, 
\end{align*}
which can be written in vector form as
\begin{equation}
	\mathbb{V}[\Palpha] = \boldsymbol{\alpha}^T \Sig \boldsymbol{\alpha}. \label{eq:varMIScov}
\end{equation}

\noindent In the following section, we provide an explicit formula to find the optimal weights $\boldsymbol{\alpha}$ for the general case of (possibly)-correlated estimators $P_1,\dots,P_k$; while in Section \ref{sec:Uncorr} we discuss the case of independent estimators, such as those constructed with the MFIS method.

\subsection{Optimizing the weights for minimum-variance estimate}\label{sec:OptWeights}
Problem~\eqref{eq:minprob} seeks the optimal $\boldsymbol{\alpha}$ such that the variance in~\eqref{eq:varMIScov} is minimized and $\Palpha$ remains unbiased. In this section, we show that such weights exist, are unique, and present a closed-form solution, provided that the covariance matrix $\Sig$ is invertible. This is summarized in the following result. 

\begin{proposition} \label{prop:weights}
	Let $\boldsymbol{P} = [P_1, \ldots, P_k]^T$ be the vector of probability estimators and assume that $\Sig$ is not singular.  Define $\boldsymbol{1}_k = [1,  \dots, 1]^T$ as a column-vector of length $k$.  The optimization problem \eqref{eq:minprob} has the unique solution
	\begin{equation*}
		\boldsymbol{\alpha} = \frac{\Sig^{-1} \boldsymbol{1}_k }{ \boldsymbol{1}^T_k \Sig^{-1} \ \boldsymbol{1}_k }.
	\end{equation*}
	That is, the minimal variance unbiased estimator $\Palpha$ is such that
	\begin{equation*}
		\Palpha = \frac{ \boldsymbol{1}^T_k \Sig^{-1} \ \boldsymbol{P}}{ \boldsymbol{1}^T_k \Sig^{-1} \ \boldsymbol{1}_k }, \qquad 
		\mathbb{V}[\Palpha] = \frac{1}{ \boldsymbol{1}^T_k \Sig^{-1} \ \boldsymbol{1}_k }.
	\end{equation*}
\end{proposition}
\begin{proof}
	We have seen above that $\sum_{i=1}^k \alpha_i=1$ if and only if  $\mathbb{E}[\Palpha] = P$. Define the cost function  $J(\boldsymbol{\alpha}) := \mathbb{V}[\Palpha] =  \boldsymbol{\alpha}^T \Sig \boldsymbol{\alpha}$ by using equation~\eqref{eq:varMIScov}. Therefore, the optimization problem~\eqref{eq:minprob} can be written as the quadratic program 
	\begin{equation}
		\min_{\boldsymbol{\alpha}} J(\boldsymbol{\alpha}) = \boldsymbol{\alpha}^T \Sig \boldsymbol{\alpha}, \quad \text{s.t.}  \ \ \boldsymbol{\alpha}^T \boldsymbol{1} =1. \label{eq:optProbAlpha}
	\end{equation}
	Letting $\mathcal{L}(\boldsymbol{\alpha},\lambda) := \boldsymbol{\alpha}^T \Sig \boldsymbol{\alpha} + \lambda (\boldsymbol{\alpha}^T \boldsymbol{1} -1) $ denote the Lagrangian cost function associated to \eqref{eq:optProbAlpha},  the optimality conditions are $\nabla_{\boldsymbol{\alpha}} \mathcal{L}(\boldsymbol{\alpha},\lambda) = \boldsymbol{0} $ and $\frac{d \mathcal{L}}{d\lambda}(\boldsymbol{\alpha},\lambda) = 0$. 
	This optimality system is written as
	\begin{equation}
		\begin{bmatrix}  \Sig & \boldsymbol{1}_k \\ \boldsymbol{1}^T_k & 0 \end{bmatrix} \begin{bmatrix} \boldsymbol{\alpha} \\ \lambda  \end{bmatrix} = \begin{bmatrix} \boldsymbol{0}_k \\ 1 \end{bmatrix}. \label{eq:QuadProg}
	\end{equation}
	For invertible $\Sig$, the unique weights to this quadratic program are then obtained by
	\begin{equation}
		\boldsymbol{\alpha} = \frac{\Sig^{-1} \boldsymbol{1} }{ \boldsymbol{1}^T_k \Sig^{-1} \ \boldsymbol{1}_k },
	\end{equation}
	and the expression for the variance follows by inserting these weights into \eqref{eq:optProbAlpha}. The estimator is obtained by inserting the weights into \eqref{eq:Palpha}. 
\end{proof}

The weights can be expressed explicitly in terms of the components of the covariance matrix as
\begin{equation}
	\alpha_i = \frac{1}{ \sigma_i^2} \left [\frac{1}{\sum_{l=1}^k \frac{1}{\sigma_l^2}}\left (  1 + \sum_{l=1}^k \frac{1}{ \sigma_l^2} \sum_{j>l}^{k} \alpha_j \sigma_l \sigma_j \rho_{l,j} \right ) - \sum_{j>i}^{k} \alpha_j \sigma_i \sigma_j \rho_{i,j}  \right ]. \label{eq:weightCompNotation}
\end{equation}	
Note, that the weights are inversely proportional to the variance of the individual estimators and the weight $\alpha_i$ depends on the covariance  between the estimators $\Pi$ and $\Pj$. Also, note that if $\Pi$ are correlated some weights may be negative, while for a diagonal $\Sig$ all weights $\alpha_i$ are \emph{strictly} positive. In the next section, we have a closer look at the uncorrelated case. 

\subsection{The special case of uncorrelated estimators} \label{sec:Uncorr}
In the situation where all estimators are uncorrelated, we recover the classical result of the \textit{inverse variance-weighted mean} \cite{meier1953varianceWeighting}.  
As a corollary from Proposition \ref{prop:weights} we get the following result. 

\begin{corollary}
	Consider the setting from Proposition \ref{prop:weights}, and let $\Sig = \textup{diag}(\sigma_1^2,\ldots, \sigma_k^2)$ be diagonal. Then the unique solution to the optimization problem \eqref{eq:minprob} is given by
	\begin{equation}
		\alpha_i = \frac{1}{\sigma_i^2  \sum_{i=1}^k \frac{1}{\sigma_i^2}}, \qquad \mathbb{V}[\Palpha] = \frac{1}{\sum_{i=1}^k \frac{1}{\sigma_i^2}} \, . \label{eq:alphaRel}
	\end{equation}
\end{corollary}

\noindent A few observations about this special case are in order:
\begin{enumerate}
	\item The optimal coefficients $\alpha_i$ of the combined estimator $\Palpha$ are inversely proportional to the asymptotic variance $\sigma_i$ of the corresponding estimator $\Pi$. To reduce the variance via a weighted combination of estimators, smaller weights are assigned to estimators with larger variance.
	\item If one variance is small compared to all other ones, say $\sigma_1^2 \ll \sigma_i^2, \ i=2,\ldots,k$, then $\sum_{i=1}^k \frac{1}{\sigma_i^2} \approx \frac{1}{\sigma_1^2}$ so that $\mathbb{V}[\Palpha] \approx \sigma_1^2$. The estimators with large variance cannot reduce the variance of the fused estimator much more. 
	\item If all estimators have equal variance, $\sigma_1^2 = \ldots = \sigma_k^2$, then $\sum_{i=1}^k \frac{1}{\sigma_i^2} = \frac{k}{\sigma_1}$ so that $\mathbb{V}[\Palpha] = \frac{\sigma_1^2}{k} $. Hence, combining the estimators reduces the variance by a factor of $k$. 
	%
	\item Since $0 < \alpha_i<1, \forall i$, it follows from both equations in \eqref{eq:alphaRel} that 
	\begin{equation}
		\mathbb{V}[\Palpha] =  \frac{1}{\sum_{i=1}^k \frac{1}{\sigma_i^2}} = {\sigma_i^2} \alpha_i < \sigma_i^2, \forall i \quad \Rightarrow \quad   \mathbb{V}[\Palpha] <  \min_{i=1,\ldots,k} \sigma_i^2. 
	\end{equation}
	Consequently, we are guaranteed to reduce the variance in $\Palpha$ by combining all estimators in the optimal way described above. 
\end{enumerate}

\subsection{Fused multifidelity importance sampling: Algorithm and Analysis}  \label{sec:fusedIS}
We now use the general fusion framework to obtain a failure probability estimator. Thus, we solve Problem~\ref{probdef} in the context of importance-sampling-based failure probability estimators so that $\Pi = P_{n_i}^{\text{IS}}$. Our proposed method optimally fuses the $k$ MFIS estimators from \eqref{eq:PniIS}, such that 
\begin{equation}
	P_{\boldsymbol{\alpha}} = \sum_{i=1}^k \alpha_i P_{n_i}^{\text{IS}},  \label{eq:PalphaIS}
\end{equation}
with the optimal weights chosen as in Proposition~\ref{prop:weights} and $\sum_{i=1}^k n_i = n$. Since estimator $P_{n_i}^{\text{IS}}$ is computed from $n_i$ samples, $\Palpha$ uses $n = \sum_{i=1}^k n_i$ samples. 

We now discuss how $\Palpha$  compares to a single importance sampling estimator with $n$ samples. Consider the estimator $P_{j'}^{\text{IS}}$ that uses $n$ samples drawn from a single biasing density $q_{j'}$ for $j' \in \{1, \ldots, k\}$. This estimator would require selection of the lowest biasing density \textit{a priori}, a formidable task. The next results compares $\Palpha$ and $P_{j'}^{\text{IS}}$, and gives a criterion for which the former has lower variance than the latter. 
\begin{proposition} \label{prop:compEstimators}
	Let $k$ estimators $P_{n_i}^{\text{IS}}$ with $n_1=n_2=\ldots=n_k$ samples be given.
	Let $j' \in \{ 1, \ldots, k\}$, and $q_{j'}$ be a biasing density that is used to derive an IS estimator $P_{j'}^{\text{IS}}$ with $n=kn_1$ samples.  If
	\begin{equation*}
		\sigma_{q_j'}^2> \frac{k}{\sum_{i=1}^k \frac{1}{\sigma_i^2}}
	\end{equation*}
	then the variance of the fused estimator $P_{\boldsymbol{\alpha}}$ in \eqref{eq:PalphaIS} with $n$ samples is smaller than the variance of the estimator with biasing density $q_{j'}$ with $n$ samples, i.e., 
	\begin{equation*}
		\mathbb{V}[\Palpha] < \mathbb{V}[P_{j'}^{\text{IS}}].
	\end{equation*}
\end{proposition}	
\begin{proof}
	Set $n_i = n/k, \ i=1, \ldots, k$, so that all estimators use the same number of samples. According to equation~\eqref{eq:alphaRel}, 
	\begin{equation*}
		\mathbb{V}[\Palpha] = \frac{1}{ \sum_{i=1}^k \frac{n_i}{\sigma_{q_i}^2}} = \frac{k}{n\sum_{i=1}^k \frac{1}{\sigma_{q_i}^2}}
	\end{equation*}
	as well as $\mathbb{V}[P_{j'}^{\text{IS}}] = \frac{\sigma^2_{q_{j'}}}{n}$, so that 
	\begin{equation*}
		\mathbb{V}[\Palpha] = \frac{k}{n \sum_{i=1}^k \frac{1}{\sigma_{q_i}^2}} < \frac{\sigma^2_{q_{j'}}}{n}  = \mathbb{V}[P_{j'}^{\text{IS}}] \qquad \Leftrightarrow  \qquad \frac{k}{\sum_{i=1}^k \frac{1}{\sigma_{q_i}^2}} < \sigma^2_{q_{j'}}
	\end{equation*}
\end{proof}

The importance sampling estimate~\eqref{eq:PniIS} requires evaluating the high-fidelity model at $n_i$ samples from the biasing density. While not required, we use $n_i = n/k, \ i=1, \ldots,k$ to distribute the computational load evenly. Extension of Proposition~\ref{prop:compEstimators} is straightforward to the case with different number of samples $n_j$ for each estimator $P_j$

The computational procedure is summarized in Algorithm~\ref{alg:MIS}. Here, we denote sampling-based estimates as $\hat{P}_{n_i}^{\text{IS}}$, which are realizations of the estimator $P_{n_i}^{\text{IS}}$.  

\begin{algorithm}[H] 
	\caption{Computing failure probability estimate $\hat{P}_{\boldsymbol{\alpha}}$ via fused importance sampling}
	\begin{algorithmic}[1] \label{alg:MIS}
		\REQUIRE Nominal distribution $p$, biasing distributions $\{ q_i \}_{i=1}^k$, \# of evaluations $\{ n_i \}_{i=1}^k$, limit state function $\limitg(\cdot)$.
		\ENSURE Failure probability estimate $\hat{P}_{\boldsymbol{\alpha}}$ and variance estimate $\mathbb{V}[\hat{P}_{\boldsymbol{\alpha}}]$
		\FOR[Loop over all surrogates] {$j=1:k$ } 
		\STATE Draw $\param_{j,1},\ldots, \param_{j,n_j}$ independent realizations from $\RV_{j}$ with density $q_j$ and compute
		\begin{equation}
			\hat{P}_{n_j}^{\text{IS}} = \frac{1}{n_j} \sum_{i=1}^{n_j} I_\failset (\param_{j,i}) \frac{p(\param_{j,i})}{q_j(\param_{j,i})} \label{eq:PniIShat}
		\end{equation}
		\STATE Compute the sample variances 
		\begin{equation}
			\hat{\sigma}_{q_j}^2 = \frac{1}{n_j - 1} \sum_{i=1}^{n_j} \left (I_\failset(\param_{j,i})\frac{p(\param_{j,i})}{q_j(\param_{j,i})} - \hat{P}_{n_j}^{\text{IS}} \right )^2  \label{eq:sampleVar}
		\end{equation}		
		\ENDFOR
		\STATE Define the vector $\boldsymbol{P} = [\hat{P}_{n_1}^{\text{IS}}, \ldots, \hat{P}_{n_k}^{\text{IS}}]^T$
		\STATE Let $\hat{\Sig} = {\textup diag}(\hat{\sigma}_{q_1}^2/n_1, \ldots, \hat{\sigma}_{q_k}^2/n_k)$
		\STATE Compute the fused estimate as in \eqref{eq:optProbAlpha}:
		\begin{equation}
			\hat{P}_{\boldsymbol{\alpha}} = \frac{ \boldsymbol{1}^T_k \hat{\Sig}^{-1} \ \boldsymbol{P}}{ \boldsymbol{1}^T_k \hat{\Sig}^{-1} \ \boldsymbol{1}_k }, \qquad 
			\mathbb{V}[\hat{P}_{\boldsymbol{\alpha}}] = \frac{1}{ \boldsymbol{1}^T_k \hat{\Sig}^{-1} \ \boldsymbol{1}_k }
		\end{equation}
	\end{algorithmic}
\end{algorithm}
%

\subsection{Error measures and practical computation} \label{sec:errComp}
The failure probability estimate $\hat{P}_{n_i}^{\text{IS}}$ is computed as in \eqref{eq:PniIShat} and the sample variance $\hat{\sigma}_{q_i}^2$ as in \eqref{eq:sampleVar}.
The root-mean-squared-error (RMSE) of the estimate $\hat{P}_{n_i}$ is
\begin{equation}
	e^{\text{RMSE}} (\hat{P}_{n_i}) = \sqrt{\frac{\hat{\sigma}_{q_i}^2}{n_i}},  \label{eq:rsme}
\end{equation}
and the relative mean-squared-error, or coefficient of variation is computed as
\begin{equation}
	e^{\text{CV}} (\hat{P}_{n_i}) = \sqrt{\frac{\hat{\sigma}_{q_i}^2}{n_i ( \hat{P}_{n_i}^{\text{IS}})^2}}. \label{eq:coefvar}
\end{equation}

\section{Test case: Convection-diffusion-reaction} \label{sec:CDR}
We first consider a PDE model whose solution can be numerically evaluated with moderate computational cost. With this model, we demonstrate the asymptotic behavior of our method because we can afford to sample the high-fidelity model $n = 10^5$ times, which will be too costly for the model in Section~\ref{sec:resultsjet}. The test problem is the convection-diffusion-reaction PDE introduced in Section~\ref{sec:CDRproblem}. Its discretizations and reduced-order models are described in Section~\ref{sec:CDRdiscrete}. Numerical results are presented in Section~\ref{sec:resultsCDR}.

\subsection{Convection-diffusion-reaction PDE model} \label{sec:CDRproblem}
\begin{figure}[H]
	\centering
	\includegraphics[width=8cm]{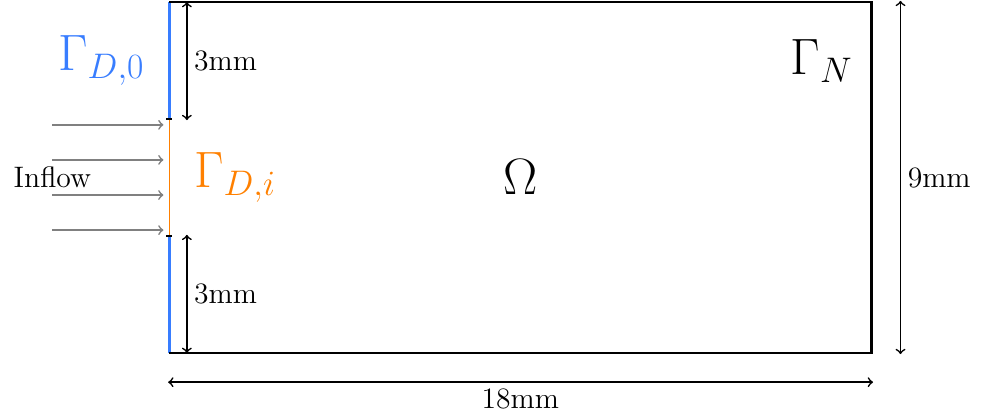}
	\caption{Set-up of combustor, with details of the boundary conditions in Table~\ref{tbl:bcs}.}
	\label{fig:combustor}
\end{figure}

We consider a simplified model of a premixed combustion flame at constant and uniform pressure, and follow the notation and set-up in \cite[Sec.3]{buffoni10MORreactingFlows}. The model includes a one-step reaction of the species
\begin{equation*}
	2H_2 + O_2 \ \rightarrow 2 H_2 O
\end{equation*}
in the presence of an additional non-reactive species, nitrogen.
The physical combustor domain is $18mm$ in length ($x$-direction), and $9mm$ in height ($y$-direction), as shown in Figure~\ref{fig:combustor}. 
\noindent The velocity field $U$ is set to be constant in the positive $x$ direction, and divergence free. The molecular diffusivity $\kappa$ is modeled as constant, equal and uniform for all species and temperature. The PDE model is given by
\begin{align}
	0 & = \kappa \Delta \state - U \nabla \state + \NL(\state,\param) \quad \in \stateSpace 
\end{align} 
where the state is comprised of the components $\state = [T, Y_{H_2}, Y_{O_2}, Y_{H_2O}] $, with the $Y_i$ being the mass fractions of the species (fuel, oxidizer, product), and $T$ denoting the temperature. 
Referring to Figure~\ref{fig:combustor}, we have that $\Gamma_D=\Gamma_{D,i} \cup \Gamma_{D,0}$ is the Dirichlet part of the boundary and $\Gamma_N$ combines the top, bottom and right boundary, where Neumann conditions are prescribed. In sum, $\partial \stateSpace = \Gamma_D \cup \Gamma_N$;  the boundary conditions are imposed as given in Table~\ref{tbl:bcs}.  
The nonlinear reaction term $\NL(\state,\param) = [\NL_T, \NL_{H_2}, \NL_{O_2}, \NL_{H_2O}](\state,\param)$ is of Arrhenius type \cite{cuenot96diffusionFlames}, and modeled as
\begin{align}
	\NL_i(\state,\param) & =   -\nu_i \left ( \frac{W_i}{\rho} \right ) \left (\frac{\rho Y_F}{W_F} \right )^{\nu_F} \left (\frac{\rho Y_O}{W_O} \right )^{\nu_O} A \exp \left ( - \frac{E}{RT} \right ), \quad i = H_2, \ O_2, \ H_2O\\
	\NL_T(\state,\param) & = Q \  \NL_{H_2O}(\state,\param).
\end{align}
The parameters of the model are defined in Table~\ref{tbl:param}. The uncertain parameters are the pre-exponential  factor $A$ and the activation energy $E$ of the Arrhenius model. The domain for these parameters is denoted as $\paramdom$. In particular, we have that 
\begin{equation*}
	\param = [A,E] \in \paramdom = [5.5 \times10^{11}, \  1.5 \times 10^{13}] \times [ 1.5 \times 10^{3}, \  9.5 \times 10^{3}].
\end{equation*} 

\begin{table}
	\centering
	\caption{Boundary conditions for the combustion model from \cite{buffoni10MORreactingFlows}.}
	\begin{tabular}{ c  l  l}
		Boundary & Temperature & Species  \\
		\hline 
		$ \Gamma_{D,i} $ & $T=950$K  & $Y_{H_2} = 0.0282, \ Y_{O_2} =0.2259, \ Y_{H_2O}=0$  \\
		$ \Gamma_{D,0} $ & $T =300$K &  $Y_{H_2} = 0, \ Y_{O_2} =0, \ Y_{H_2O}=0$  \\
		$ \Gamma_N $ & $\nabla T \cdot \boldsymbol{n} = 0$ &  $\nabla Y_i\cdot \boldsymbol{n} = 0 $  \\
	\end{tabular}
	\label{tbl:bcs}
\end{table}

\begin{table}
	\centering
	\caption{Parameters for the combustion model from \cite{buffoni10MORreactingFlows}.}
	\begin{tabular}{c  l  l   l}
		quantity & physical meaning & assumptions & value  \\
		\hline 
		$\kappa$ &  molecular diffusivity & const., equal, uniform $\forall i$ & $2.0\frac{\text{cm}^2}{\text{s}}$ \\
		$U$      & velocity               & const.    & $50 \frac{\text{cm}}{\text{s}} $ \\
		$W_{H_2}$& molecular weight       & const.    &  $2.016 \frac{\text{g}}{\text{mol}}$ \\
		$W_{O_2}$& molecular weight       & const.    & $  31.9\frac{\text{g}}{\text{mol}}$ \\
		$W_{H_2 O}$& molecular weight     & const.    & $ 18 \frac{\text{g}}{\text{mol}}$ \\
		$ \rho $ & density of mixture     & const.    & $1.39 \times 10^{-3} \frac{\text{g}}{\text{cm}^3}$ \\
		$R$      & univ. gas constant     & const.    & $8.314472 \frac{\text{J}}{\text{mol} \ \text{K}}$\\
		$Q$      & heat of reaction       & const.    & $9800$K \\
		$\nu_{H_2}$ & stochiometric coefficient & const. & 2 \\
		$\nu_{O_2}$ & stochiometric coefficient & const. & 1 \\
		$\nu_{H_2O}$ & stochiometric coefficient & const. & 2 \\
	\end{tabular}
	\label{tbl:param}
\end{table}

\subsection{Discretization and reduced-order models} \label{sec:CDRdiscrete}
The model is discretized using a finite difference approximation in two spatial dimensions, with 72 nodes in $x$ direction, and 36 nodes in $y$ direction, leading to $10,804$ unknowns in the model.
The nonlinear system is solved with Newton's method. Let $\mathbf{T}(\param)$ be the vector with components corresponding to the approximations of the temperature $T(x,y; \param)$ at the grid points. The high-fidelity model (HFM) is $f: \paramdom \mapsto \mathbb{R}$ and the quantity of interest is the maximum temperature over all grid points:
\begin{equation*}
	f(\param) = \max \  \mathbf{T}(\param).
\end{equation*}

Reduced-order models provide a powerful framework to obtain surrogates for expensive-to-evaluate models. In the case of nonlinear systems, reduced-order models can be obtained via reduced-basis methods \cite{RBbook16}, dynamic mode decomposition \cite{DMDbook16}, proper orthogonal decomposition \cite{berkooz1993proper}, and many others; for a survey, see \cite{MORbook17}.
Here, we compute reduced-order models $f^{(i)}$ for our multifidelity approach via Proper Orthogonal Decomposition and the Discrete Empirical Interpolation Method (DEIM) for an efficient evaluation of the nonlinear term. The training snapshots are generated from solutions to the high-fidelity model on a parameter grid of $50\times50$ equally spaced values $\param \in  \paramdom$.
The three surrogate models are built from $2,10,15$ POD basis vectors, and accordingly $2,5,10$ DEIM interpolation points. The corresponding models are denoted as ROM1, ROM2, ROM3, respectively. We denote by $\mathbf{T}_r^{(i)}(\param)$ the approximation to the temperature $T(x,y; \param)$ via the $i$th ROM.
The surrogate models $f^{(i)}$ are the mappings $f^{(i)}: \paramdom \mapsto \mathbb{R}$ with corresponding quantity of interest denoted as
\begin{equation*}
	f^{(i)}(\param) = \max \  \mathbf{T}_r^{(i)}(\param), \quad i=1,\ldots,k.
\end{equation*}
We refer the reader to \cite{buffoni10MORreactingFlows} for more details on the discretization and ROM construction for this convection-diffusion-reaction model.

\subsection{Results for multifidelity fusion of failure probabilities} \label{sec:resultsCDR}
We define a failure of the system when the maximum temperature in the combustor exceeds  $2430$K, so that the limit state function is
\begin{equation}
	g(f(\param)) = 2430 - f(\param),
\end{equation}
and likewise for the reduced-order models $g(f^{(i)}(\param)) = 2430 - f^{(i)}(\param)$.

To compute the biasing densities, we draw $\hat{m}=20,000$ samples from the uniform distribution on $\paramdom$, compute surrogate-based solutions, and evaluate the limit state function for those solutions.  If the limit state function indicates failure of the system for a solution obtained from the $i$th surrogate model, the corresponding parameter is added to $\mathcal{G}^{(i)}$, the failure set computed from the $i$th surrogate model. We compute the biasing densities $q_1,q_2,q_3$ via MFIS (see Section \ref{sec:MFIS}) as Gaussian mixture distributions with a single component. Table~\ref{tbl:timeCDR} shows the computational cost in CPU time of computing the biasing distributions from the various ROMs and the HFM. Computing a biasing density using the high-fidelity model with $\hat{m}=20,000$ samples costs approximately $2.1$ CPU-hours. Constructing the biasing density via the low-fidelity models ROM2 and ROM3 reduces the computational time by a factor of 66 and 58, respectively. Note, that ROM1 is the reduced-order model that is cheapest to execute per model evaluation, but it is also the least accurate. In our case, ROM1 did not produce any samples in the failure region, even after $\hat{m}=10^5$ samples. It is not unexpected that ROM1 is so inaccurate, since only two POD modes are not enough to resolve the important character of this problem. ROM1 is included to demonstrate how the fusion approach can be effective even in the presence of highly inaccurate surrogate models.

\begin{table}[!ht]
	\centering
	\caption{CPU time to generate the biasing densities, and the number of samples in the failure domain.  }
	\begin{tabular}{ l |  l  l l l }
		& ROM1 	  & ROM2 	& ROM3 	 & HFM  \\
		\hline 
		\# of samples drawn 								& $10^5$  &  $2\times 10^4$ & $2\times 10^4$   & $2\times 10^4$   \\
		\# of samples in failure domain $\mathcal{G}^{(i)}$ & 0 	  &  13 	& 17 	   & 17  		 \\
		time needed 										& N.A. 	  & $11.2$[s] & $12.7$[s]& $2.1$[h] \\
	\end{tabular}
	\label{tbl:timeCDR}
\end{table}

\begin{figure}[H]
	\begin{centering}
		\includegraphics[width=0.48\textwidth]{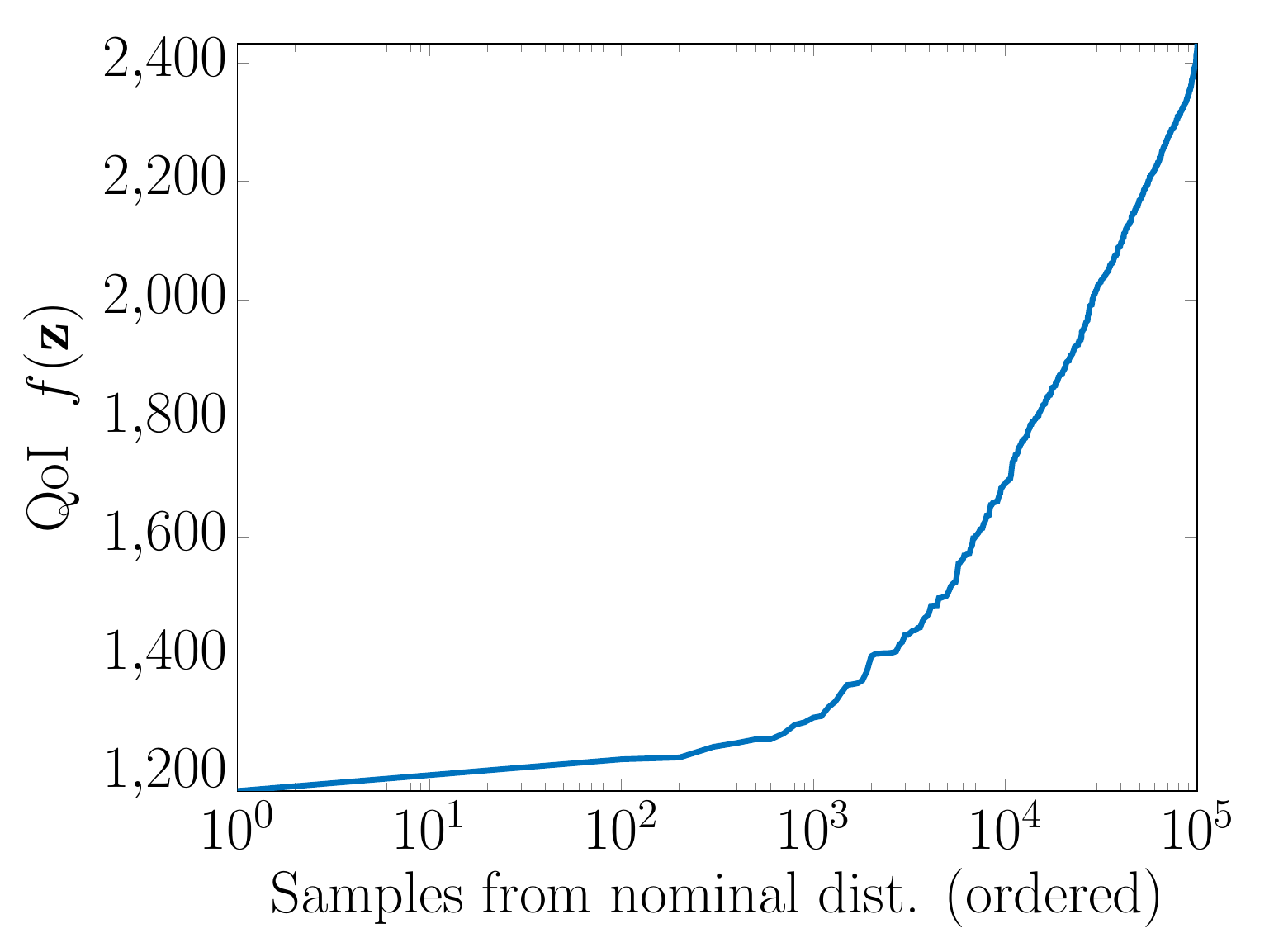}
		\hspace{0.3cm}
		\includegraphics[width=0.48\textwidth]{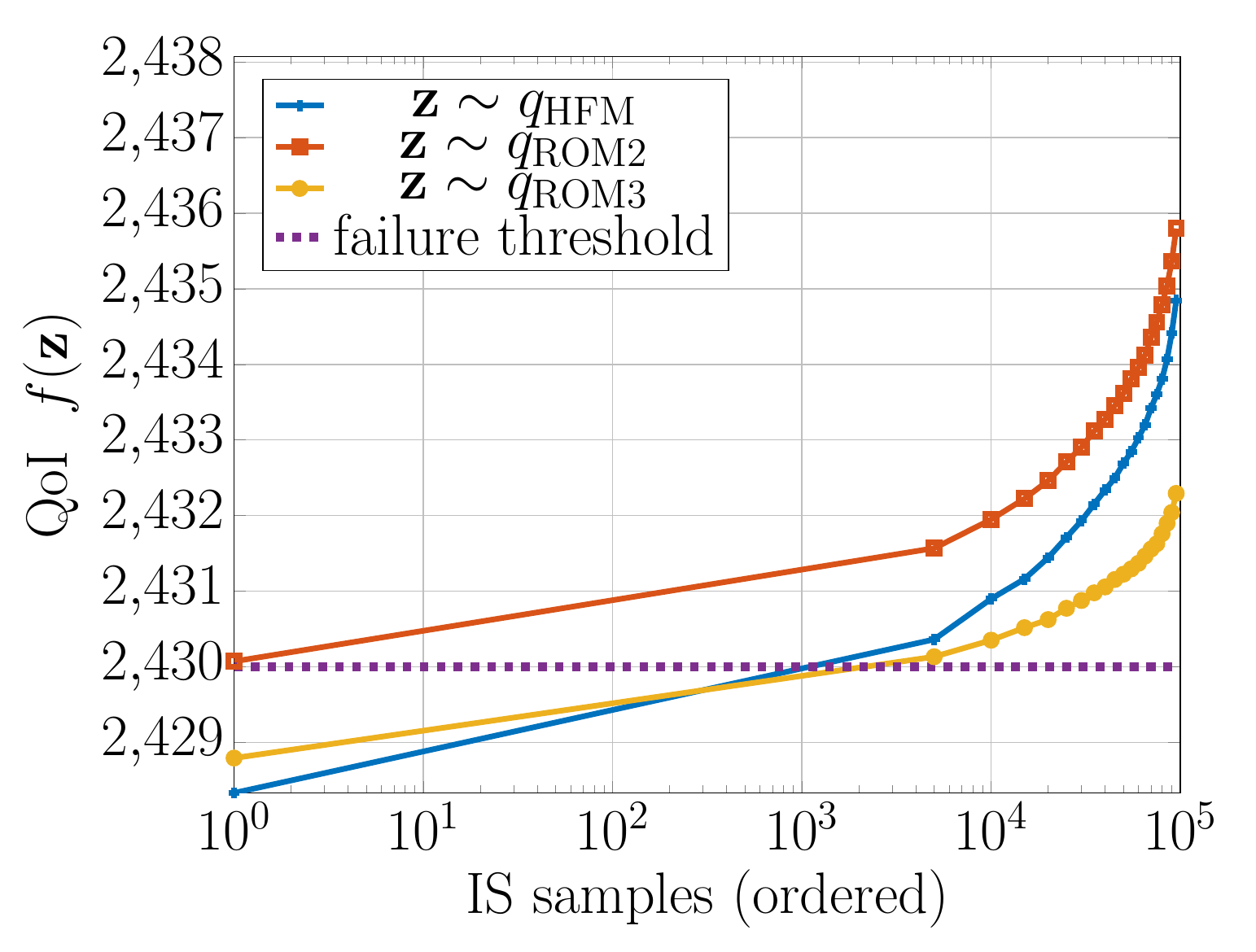}
		\caption{Quantity of interest $f(\param)$ in [K] of HFM ordered by magnitude versus \# of samples $\param$, for $m=10^5$ samples. Left: Samples are from the nominal (uniform) distribution. Right: The parameter samples are drawn from different biasing distributions (biased towards failure above $2430$K). This demonstrates that the biasing distributions are good since the outputs are largely above the failure threshold. Here, ROM1 did not have any parameters in the failure domain, and hence defaulted to being the nominal distribution and is therefore not plotted.   }
		\label{fig:CDR_QoI}
	\end{centering}
\end{figure}

In Figure~\ref{fig:CDR_QoI} we show the quantity of interest, i.e., the maximum temperature. The plots are obtained by generating $m=10^5$ samples from the nominal distribution (left) and the respective biasing distributions (right), and evaluating the HFM at those samples. Figure~\ref{fig:CDR_QoI}, left, shows that the typical range of the quantity of interest is between approximately 1200K and 2440K. However, only the events where the quantity of interest is above 2430K are relevant for the failure probability computation.   By using the biasing distributions in Figure~\ref{fig:CDR_QoI}, right, a large portion of the outputs leads to a failure of the system. This indicates that the biasing distributions are successful in generating samples at the failure region of the high-fidelity model.

Next, we show results for the fused multifidelity estimator $\Palpha$ with $n$ samples and compare it with importance sampling estimators $\hat{P}_{n_i}^{\text{IS}}$ that only use a single biasing density and also $n$ samples. The fused estimator is obtained via Algorithm~\ref{alg:MIS} with $n_i=\lfloor n/3 \rfloor, i=1,2,3,$ samples by fusing the three surrogate-model-based importance sampling estimators.
For reference purposes, a biasing density is constructed as described above using the HFM with $\hat{m}=20,000$ samples. Based on this density, we compute an importance sampling estimate of the failure probability with $n=10^5$ samples, resulting in $\hat{P}_{10^5}^{\text{IS}} = 8.42 \times 10^{-4}$.

\begin{figure}[H]
	\begin{centering}
		\includegraphics[width=0.495\textwidth]{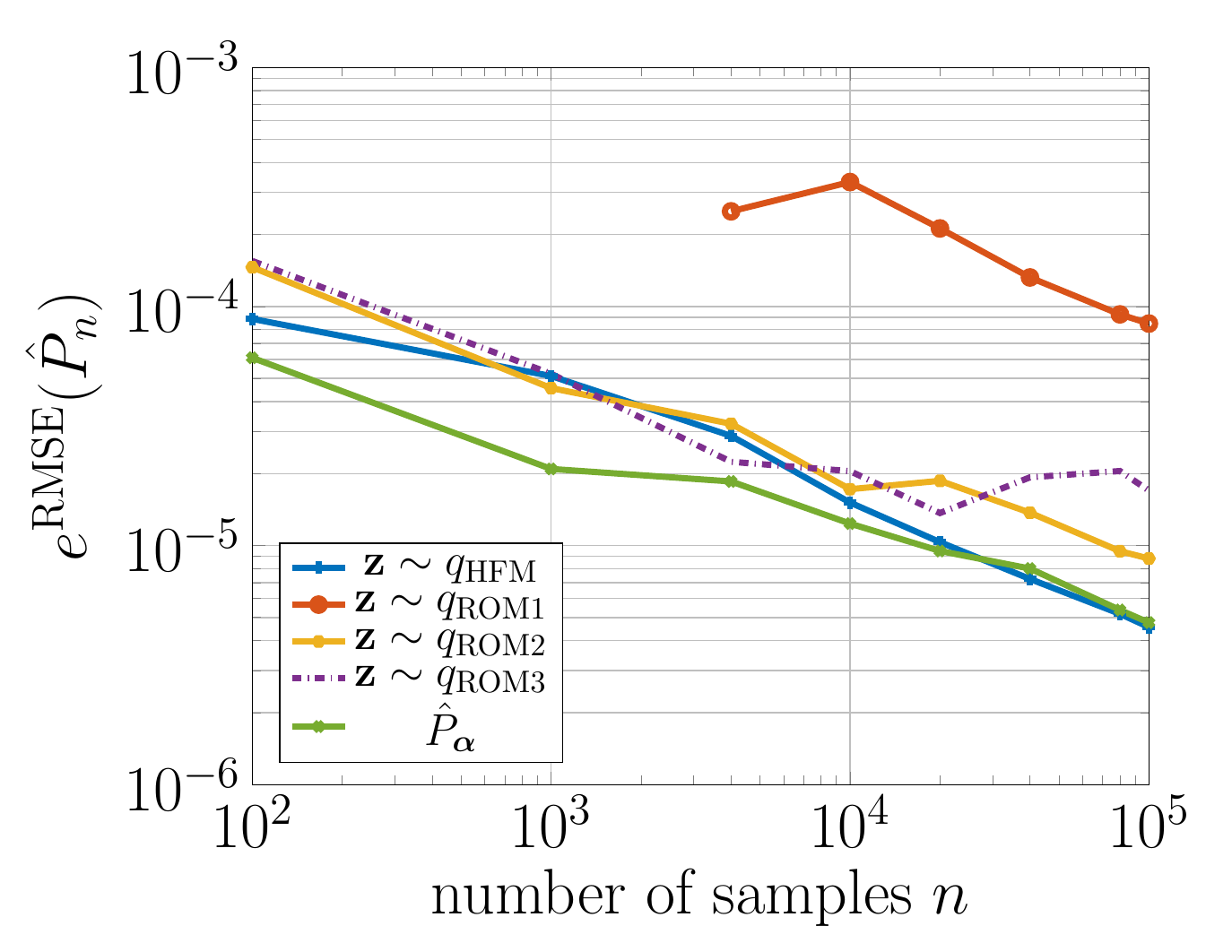}
		\includegraphics[width=0.495\textwidth]{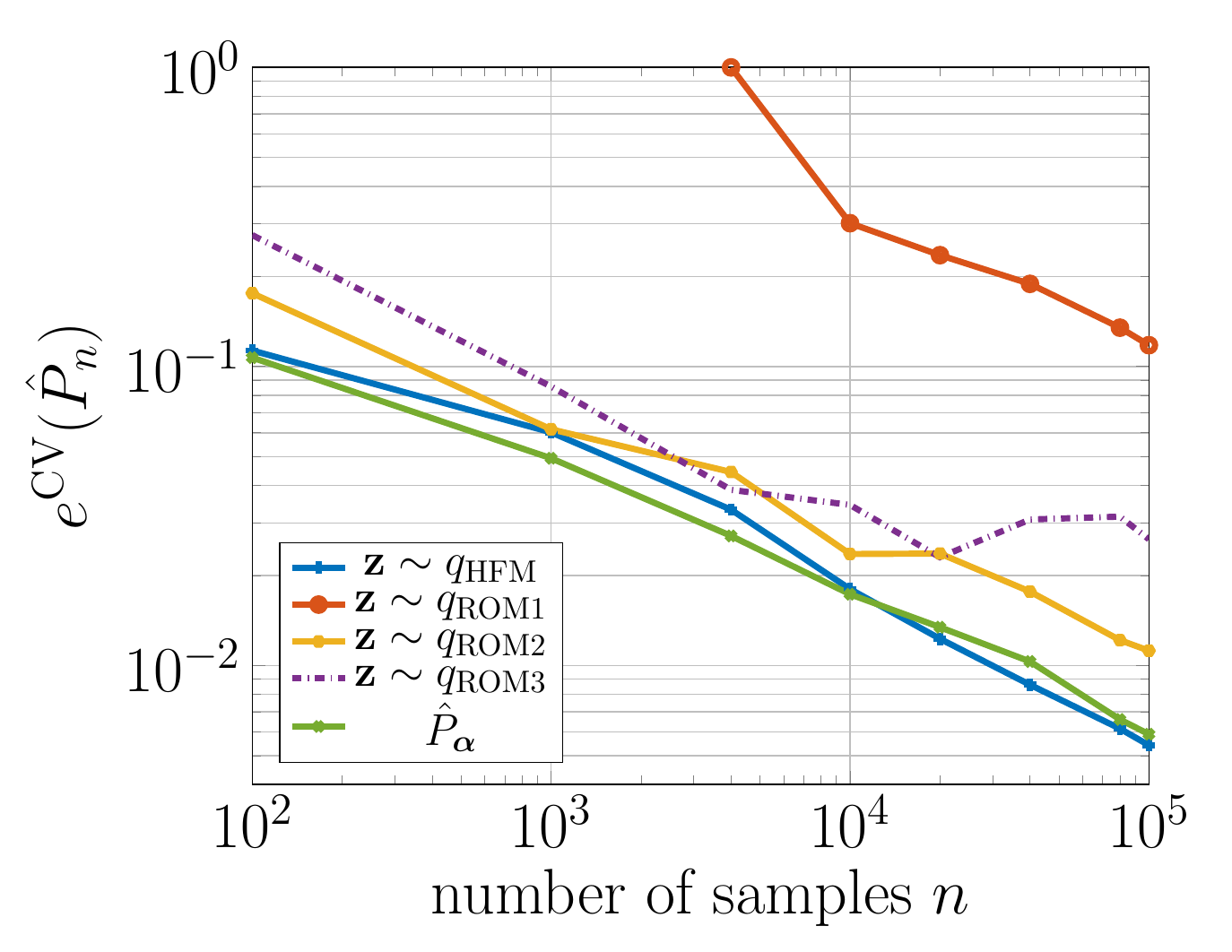}
		\caption{Left: Root mean-squared error from \eqref{eq:rsme}; Right: Coefficient of variation as defined in \eqref{eq:coefvar} for the convection-diffusion-reaction simulation. }
		\label{fig:CDR_RSME}
	\end{centering}
\end{figure}

To assess the quality of the fused estimator $\Palpha$, we consider the error measures introduced in Section~\ref{sec:errComp}. In Figure~\ref{fig:CDR_RSME}, left, we show the root mean-squared error of the importance sampling estimators $\hat{P}_{n_i}^{\text{IS}}$ as well as the combined estimator $\Phatalpha$.  Figure~\ref{fig:CDR_RSME}, right, shows the coefficient of variation defined in \eqref{eq:coefvar} for the estimators. The fused estimator is competitive in RMSE and coefficient of variation with the estimator using the high-fidelity biasing density, but comes at a much lower computational cost. 

Note, that the fused estimator does not use any of the high-fidelity information. We only plotted the high-fidelity estimator for comparison reasons, but the high-fidelity density is not used in our algorithm. Heuristically, we could expect the fused estimator to perform better than the MFIS estimator with high-fidelity-derived biasing density in the following situation. Let the HFM be so expensive that the HF biasing density is built only from a few failure samples, and assume the low-fidelity models are good surrogates, hence able to cheaply explore the failure region. Then the low-fidelity biasing density could be better than the high-fidelity biasing density.  

In Table~\ref{tbl:alpha_CDR} we show the weights for the fused estimator $\Phatalpha$. The fused estimator assigns only a small weight $\alpha_1$ to the estimator $\hat{P}_{n_1}^{\text{IS}}$ which uses biasing density $q_1$. This was expected, as the estimator has large variance due to the fact that biasing density $q_1$ is actually the nominal density, see Table~\ref{tbl:timeCDR} as the ROM1 evaluation did not yield any samples in the failure domain. 
\begin{table}[H]
	\centering
	\caption{Weights of the fused estimator $\hat{P}_{\boldsymbol{\alpha}}$ with $n$ samples.}
	\begin{tabular}{l | c c c c c c }
		& $n=10^2$  & $n=10^3$  & $n=10^4$ & $n=2\times 10^4$ & $n=4\times 10^4$  & $n=10^5$ \\
		\hline
		$\alpha_1$ & 0      &      0  & 0.005 & 0.001 & 0.002  & 0.005 \\
		$\alpha_2$ & 0.587  &  0.471  & 0.331 & 0.294 & 0.415  & 0.742 \\
		$\alpha_3$ & 0.413  &  0.529  & 0.664 & 0.705 & 0.583  & 0.253 \\
	\end{tabular}
	\label{tbl:alpha_CDR}
\end{table}

	\subsection{Comparison to subset simulation methods}	
	To demonstrate the efficiency of our proposed multifidelity method compared to state-of-the-art existing methods in failure probability estimation, we compare our results to subset simulation~\cite{AuBeck_2001SS}, a widely used method for reliability analysis and failure probability estimation. The method defines intermediate failure events 
	\begin{equation*}
		\failset_j:= \{ \param \in \paramdom \  | \ \limitg(f(\param))<b_j \}, \ j=1, \ldots, L,
	\end{equation*}
	for a sequence of threshold levels $b_1 > b_2 > \ldots >b_L=0$ and $L$ being the final level. This ensures that the intermediate failure events are nested as $\failset_1 \supset \failset_2 \supset\ldots \supset \failset_L = \failset$. The failure probability can then be expressed as
	$$
	P = P(I_\failset) = \mathbb{E}(I_{\failset_1}) \prod_{j=2}^{L} P(I_{\failset_j} \vert I_{\failset_{j-1}}).
	$$
	Thus, this method requires sampling from the conditional events $\failset_j \vert \failset_{j-1}$, and the efficiency of this sampling is pivotal to the success of subset simulation. Markov Chain Monte Carlo (MCMC) methods provide efficient solutions to this problem~\cite{papaioannou2015mcmcSS}. Note, that the $b_j$ cannot be determined in advance, but are found adaptively by specifying an intermediate failure probability $p_0 = P(\failset_j \vert \failset_{j-1})$. A typical choice is $p_0 = 0.1$ which yields efficient subset simulation results, see~\cite{AuBeck_2001SS}.
	
	Here, we compare our fused importance sampling approach for failure probability estimation to a direct application of subset simulation to the full model. We follow the recent MCMC implementation for subset simulation of \cite{papaioannou2015mcmcSS}. Table~\ref{tbl:SuS} lists the computational results that include the number of levels $L$ that subset simulation needed to arrive at the failure probability estimate, the samples at each level (user defined), the failure probability estimate, and the overall number of samples needed (not known beforehand). All results were averaged over ten independent runs. We also give an approximate coefficient of variation, although we caution that this is \textit{not} the same coefficient of variation defined in \eqref{eq:CVwithP}, since at the intermediate levels, subset simulation produces correlated samples. Thus, we used an approximated  coefficient of variation as suggested in ~\cite[Eq.~(19)]{zuev2012bayesianSS}. For a thorough discussion of the coefficient of variation estimation within subset simulation we refer to~\cite[Sec.5.3]{betz2017bayesian}. We observe from Table~\ref{tbl:SuS} that the coefficient of variation monotonically decreases as more samples are added. 	
	To compare our proposed multifidelity fusion method with subset simulation, we first note that the estimate from subset simulation $\hat{P}_f$ is biased for finite $N$ (see~\cite[Sec.6.3]{AuBeck_2001SS}), whereas our fused estimator $\hat{P}_\alpha$ is unbiased. Moreover, the numerical results in Table~\ref{tbl:SuS} show that the estimated coefficients of variation are about one order of magnitude larger than the coefficients of variation we reported in Figure~\ref{fig:CDR_RSME}, right. From a computational cost perspective, the estimator with 20,000 samples in subset simulation produces an approximated coefficient of variation of $1.18\times 10^{-1}$ whereas our fused estimator $\hat{P}_{\boldsymbol{\alpha}}$ produces a coefficient of variation of $1.34 \times 10^{-2}$ for the same number of high-fidelity model evaluations. Thus, the fused estimator outperforms subset simulation in this particular example. 
	In sum, our method can successfully take advantage of cheaper low-fidelity methods to get accurate estimators, while the subset simulation method works directly with the full model and therefore does not have access to cheaper surrogate model information. 
	\begin{table}[H]
		\centering
			\caption{Results for subset simulation to compute failure probabilities for the convection-diffusion-reaction problem.}
			\begin{tabular}{c c c c c }
				samples  & samples each level & No of levels $L$ & failure Prob.  & estimated C.o.V.  \\
				\hline
				2000  & 500  & 4 & $1.06\times 10^{-3}$ & $3.24\times 10^{-1}$ \\	
				4000  & 800  & 5 & $8.14\times 10^{-4}$ & $2.69\times 10^{-1}$ \\
				4000  & 1000 & 4 & $8.80\times 10^{-4}$ & $2.29\times 10^{-1}$ \\		
				6000  & 1500 & 4 & $9.30\times 10^{-4}$ & $1.92\times 10^{-1}$ \\	
				10000 & 2000 & 5 & $7.70\times 10^{-4}$ & $1.68\times 10^{-1}$  \\ 
				20000 & 4000 & 5 & $8.22\times 10^{-4}$ & $1.18\times 10^{-1}$  \\			
			\end{tabular}
			\label{tbl:SuS}
	\end{table}

\section{Failure probability estimation related to a free plane jet} \label{sec:resultsjet}
We apply the proposed fusion of estimators to quantify the influence of uncertain parameters on the amount of turbulent mixing produced by a free plane jet.

This is a challenging problem, since it involves an expensive-to-evaluate model for which the naive computation of low probabilities requires thousands of hours of computation. We reduce this number significantly with our multifidelity importance sampling framework via fusion of estimators.

The remainder of this section is organized as follows. Section \ref{sec:jetintro} introduces the free plane jet, followed by details of the model and its governing equations in Section~\ref{sub:jet_model}. The uncertain parameters and quantity of interest are defined in Section~\ref{sub:jet_uncertain}. The low-fidelity surrogate models used in this investigation are discussed in Section~\ref{sub:jet_surrogate}. Finally, the results for multifidelity fusion of small probability estimators are presented in Section~\ref{sub:jet_results}.

\subsection{Large-scale application: Free plane jet} \label{sec:jetintro}
Free turbulent jets are prototypical flows believed to represent the dynamics in many engineering applications, such as combustion and propulsion. As such, free jet flows are the subject of several experimental \cite{Gutmark76_planeJet, Gutmark78_impJet, Krothapalli81_mixing}
and numerical investigations \cite{Zhou99_RANS, Ribault99, Stanley02_mixing, Klein03_DNSplanejet, klein15} and constitute an important benchmark for turbulent flows.

\begin{figure}[!ht]
	\begin{center}
		\subfigure[Contours of turbulent kinetic energy.]{\includegraphics[width=0.49\textwidth]{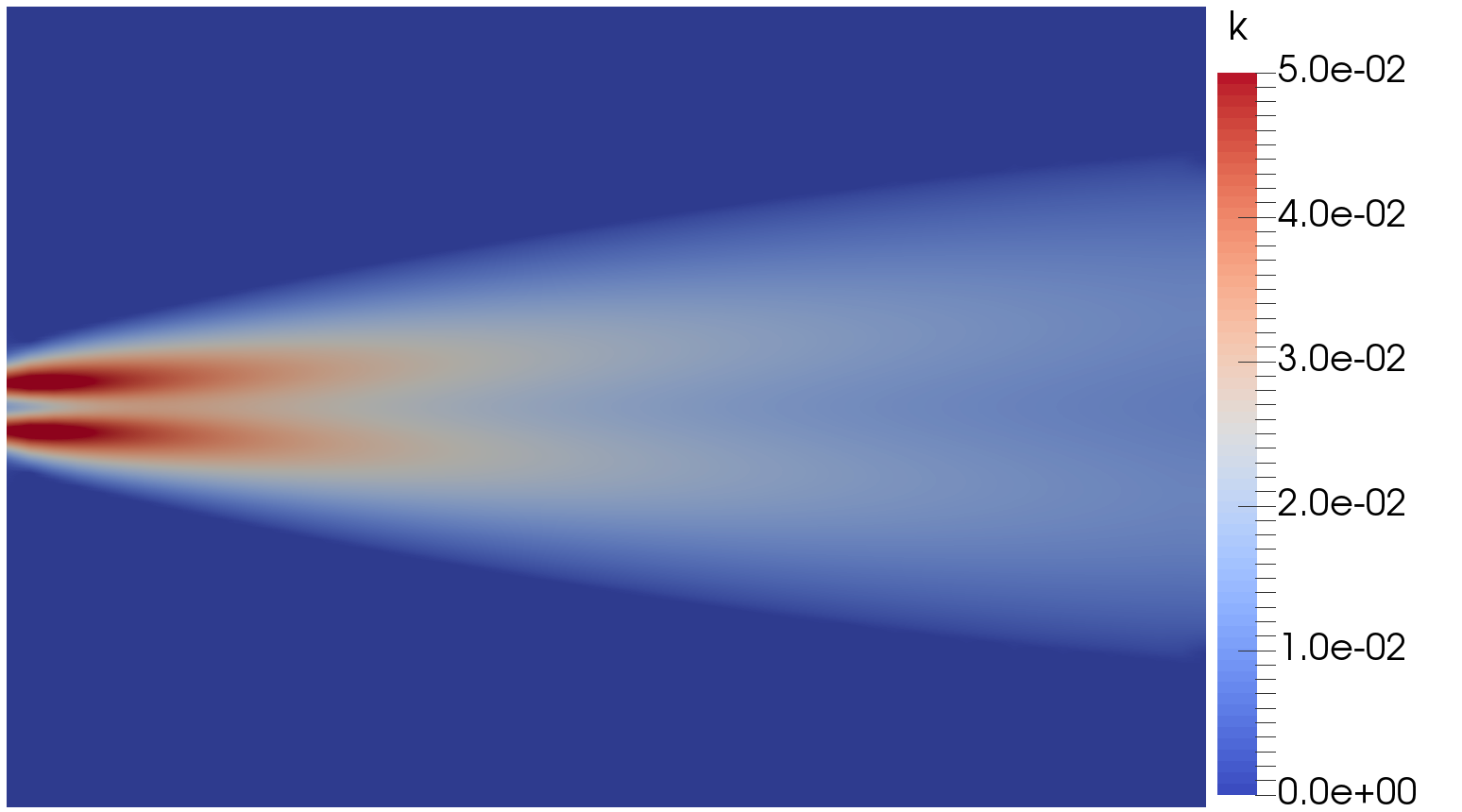}}
		\subfigure[Streamlines colored by the intensity of the velocity.]{\includegraphics[width=0.50\textwidth]{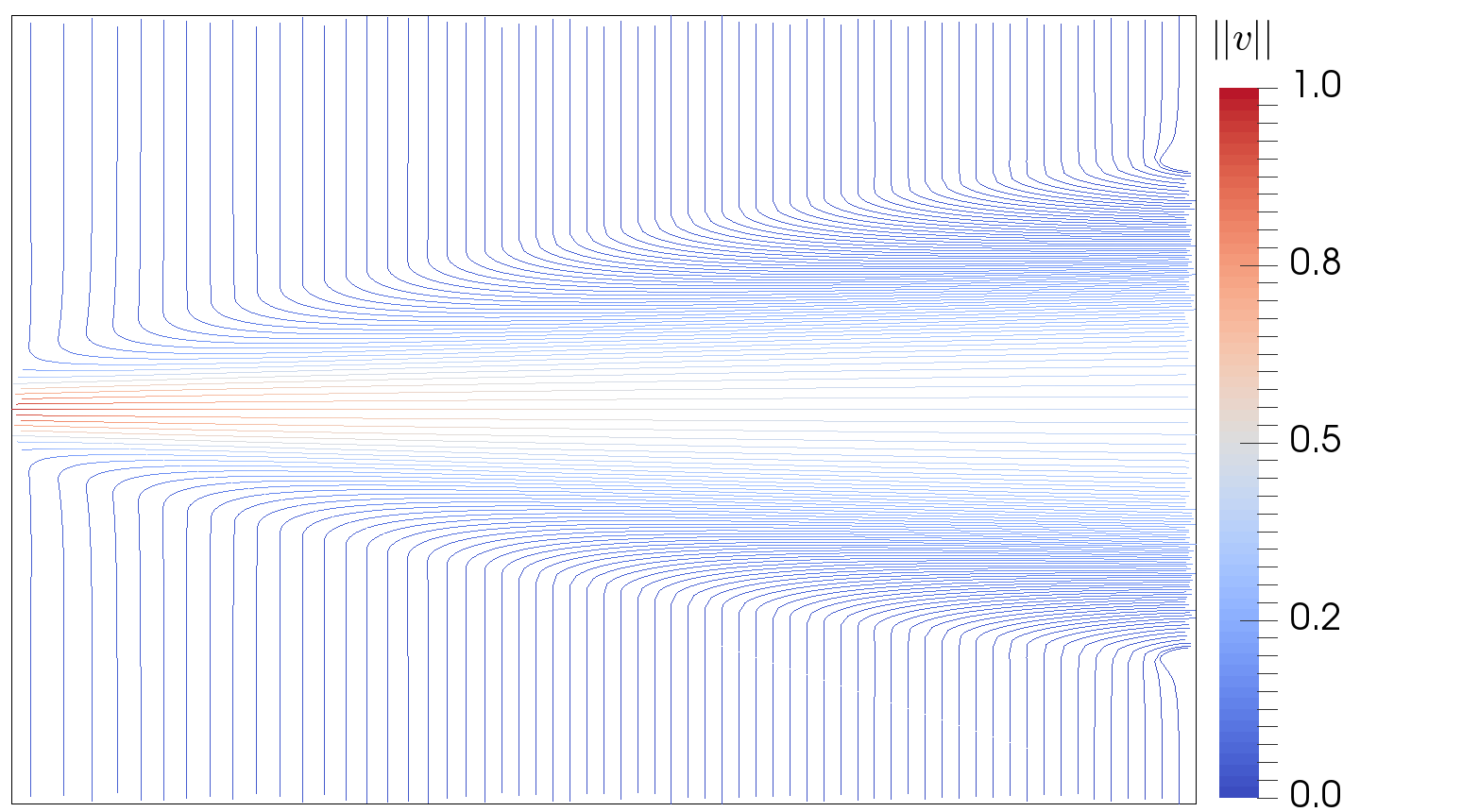}}
	\end{center}
	\caption{Flow field of a two-dimensional plane jet at Reynolds number 10,000, computed with standard coefficients of $k-\epsilon$ turbulence model.}
	\label{fig:jet_description}
\end{figure}

Our expensive-to-evaluate model of a free plane jet is based on the two-dimensional incompressible Reynolds-averaged Navier-Stokes (RANS) equations, complemented by the $k-\epsilon$ turbulence model.
Although a RANS model does not resolve all relevant turbulent features of the flow, it represents a challenging large-scale application for the computation of small probabilities. We use this model to investigate the influence of five uncertain parameters on the amount of turbulent mixing produced by the jet. We quantify turbulent mixing using a relatively simple metric: the integral jet width. One of the uncertain parameters is the Reynolds number at the inlet of the jet, which is assumed to vary from 5,000 to 15,000. The other four uncertain parameters correspond to coefficients of the $k-\epsilon$ turbulence model and its boundary condition, as detailed in Section~\ref{sub:jet_uncertain}. Figure~\ref{fig:jet_description} shows a flow field typical of the cases considered here.

\subsection{Modeling and governing equations} \label{sub:jet_model}

We consider a free plane jet in conditions similar to the ones reported in \cite{Klein03_DNSplanejet, klein15}.
Namely, the flow exits a rectangular nozzle into quiescent surroundings with a prescribed top-hat velocity profile and turbulence intensity.
The nozzle has width $D$, and is infinite along the span-wise direction.
The main difference between the free plane jet we considered here and the one described in \cite{Klein03_DNSplanejet, klein15} is the Reynolds number at the exit nozzle.
Here the Reynolds number varies between 5,000 and 15,000.

\begin{figure}[h!]
	\begin{center}
		\includegraphics[width=3.5in]{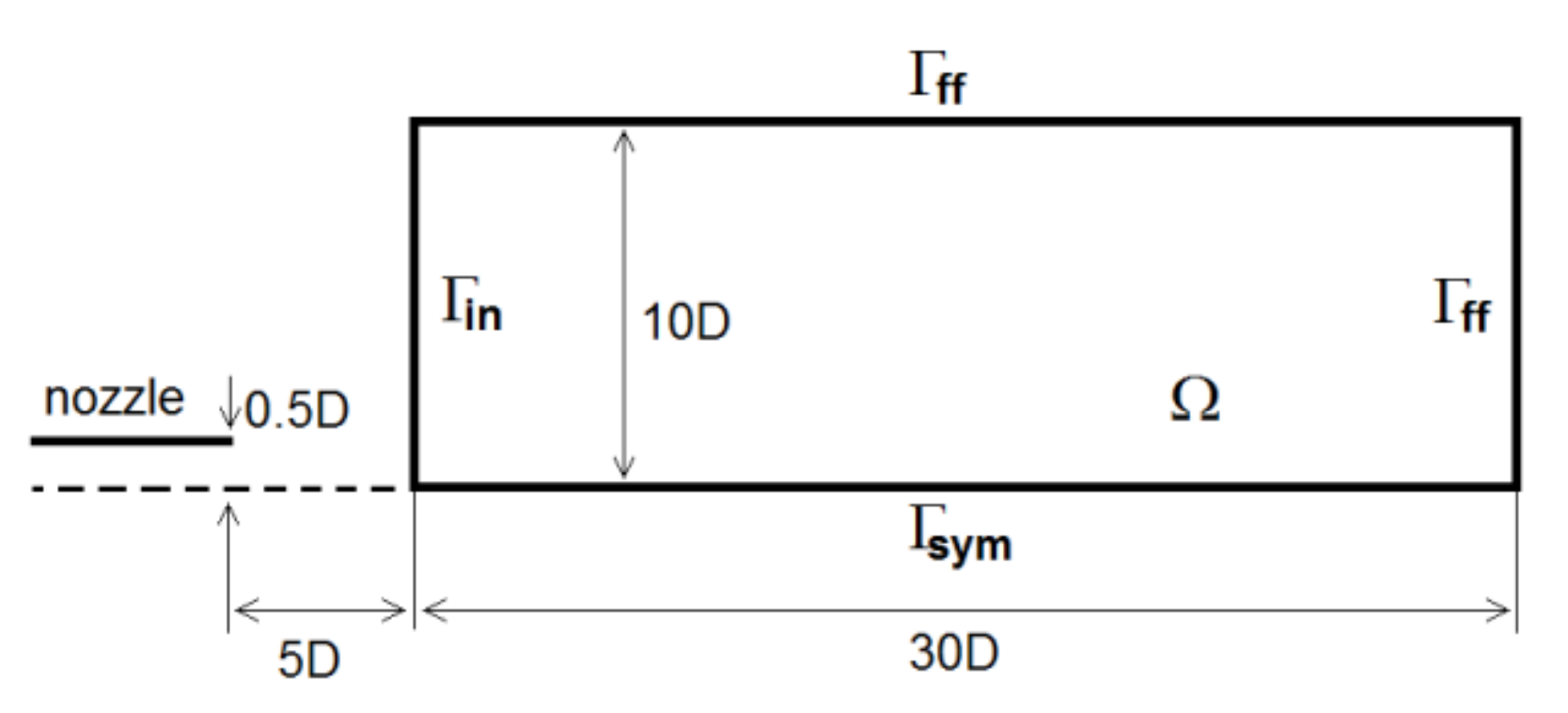}
	\end{center}
	\caption{Illustration of the free plane jet setup.
		The diameter of the nozzle is denoted by $D$. The simulation domain $\Omega$ is composed of a $30D \times 10D$ box situated at a distance $5D$ downstream to the nozzle exit.}
	\label{fig:jet_setup}
\end{figure}

Our simulation model computes the flow in a rectangular domain $\Omega$ located at a distance $5D$ downstream from the exit of the jet nozzle, as illustrated in Figure~\ref{fig:jet_setup}.
By doing so, modeling the conditions at the exit plane of the jet nozzle is avoided.
Instead, direct numerical simulation data are used to define inlet conditions at the surface $\Gamma_{\text{in}}$.
The dynamics are modeled with the incompressible Reynolds-averaged Navier-Stokes equations, complemented by the $k-\epsilon$ turbulence model \cite{Launder74_ke}:
\begin{align}
	(\boldsymbol{v} \cdot \nabla)\boldsymbol{v} 
	+ \dfrac{1}{\rho}\nabla p 
	- \nabla \cdot ((\nu + \nu_t)\bar{\bar{S}}(\boldsymbol{v}))
	&= 0,
	\label{eq:jet_mom}\\
	\nabla \cdot \boldsymbol{v} &= 0,
	\label{eq:jet_mass}\\
	\boldsymbol{v} \cdot \nabla k 
	- 2\nu_t(\bar{\bar{S}}(\boldsymbol{v}):\bar{\bar{S}}(\boldsymbol{v}))
	+ \epsilon 
	- \nabla \cdot \left(\left(\nu + \dfrac{\nu_t}{\sigma_k}\right)\nabla k \right)
	&= 0,
	\label{eq:jet_k}\\
	\boldsymbol{v} \cdot \nabla \epsilon
	- 2C_{1\epsilon}\dfrac{\epsilon\nu_t}{k}(\bar{\bar{S}}(\boldsymbol{v}):\bar{\bar{S}}(\boldsymbol{v})) 
	+ C_{2\epsilon}\dfrac{\epsilon^2}{k} 
	- \nabla \cdot \left(\left(\nu + \dfrac{\nu_t}{\sigma_{\epsilon}}\right) \nabla \epsilon \right)
	&= 0,
	\label{eq:jet_e}
\end{align}
where $\boldsymbol{v} = [v_x, v_y]$ denotes the velocity vector, $p$ denotes pressure, $\rho$ is the density, $\nu$ is the kinematic viscosity, and $\bar{\bar{S}}$ is the strain rate tensor given by
\begin{equation*}
	\bar{\bar{S}}(\boldsymbol{v}) = \dfrac{1}{2}(\nabla \boldsymbol{v} + (\nabla \boldsymbol{v})^T).
\end{equation*}
	In the $k-\epsilon$ turbulence model, $k$ denotes the turbulent kinetic energy, $\epsilon$ denotes the turbulent dissipation, and $\nu_t$ denotes the turbulent kinematic viscosity, defined as
	\begin{equation}
		\nu_t = C_{\mu}\dfrac{k^2}{\epsilon} \label{eq:nu_t}.
	\end{equation}
	The coefficients\footnote{We use $\sigma_k$ and $\sigma_\epsilon$ here as model coefficients, which is typical notation in fluids community. These are only used in this section, and throughout the paper $\sigma$'s are variances.} $C_{\mu}$, $C_{1\epsilon}$, $C_{2\epsilon}$, $\sigma_k$, $\sigma_{\epsilon}$ in \eqref{eq:jet_k}--\eqref{eq:nu_t} are either considered as uncertain parameters, or are functions of uncertain parameters, as detailed in Section~\ref{sub:jet_uncertain}.

At the inlet surface $\Gamma_{\text{in}}$ Dirichlet boundary conditions are imposed.
Data obtained by the direct numerical simulation described in \cite{klein15} (Reynolds number 10,000) are used to determine reference inlet profiles for velocity, $\boldsymbol{v_{\text{ref}}}$, and for turbulent kinetic energy, $k_{\text{ref}}$.
Inlet conditions are allowed to vary by defining a velocity intensity ($U$) scale, which is applied to the reference profiles.
Turbulent dissipation at the inlet is estimated by assuming a mixing length model.
Thus, the boundary conditions at the inlet surface are given by
\begin{align*}
	\boldsymbol{v}|_{\Gamma_{\text{in}}} &= U\boldsymbol{v_{\text{ref}}}, &
	k|_{\Gamma_{\text{in}}} &= U^2k_{\text{ref}}, &
	\epsilon|_{\Gamma_{\text{in}}} = C_{\mu}\dfrac{k^{3/2}}{\ell_m},
\end{align*}
where $\ell_m$ denotes the mixing length parameter.

At the symmetry axis surface, $\Gamma_{\text{sym}}$, no-flux boundary conditions are imposed through a combination of Dirichlet and Neumann conditions of the form
\begin{align*}
	v_y|_{\Gamma_{\text{sym}}} &= 0, &
	\left.
	\dfrac{\partial v_x}{\partial n}
	\right|_{\Gamma_{\text{sym}}} &= 0, &
	\left.
	\dfrac{\partial k}{\partial n}
	\right|_{\Gamma_{\text{sym}}} &= 0, &
	\left.
	\dfrac{\partial \epsilon}{\partial n}
	\right|_{\Gamma_{\text{sym}}} &= 0.
\end{align*}
Finally, at the surface $\Gamma_{\text{ff}}$ ``far-field'' conditions that allow the entrainment of air around the jet are imposed through weak Dirichlet conditions, as detailed in \cite{Villa17}.

The complete model includes additional features that make it more amenable to numerical discretization.
The most delicate issue in the solution of the RANS model is the possible loss of positivity of the turbulence variables.
To avoid this issue, we introduce an appropriately mollified (and thus smoothly differentiable) max function to ensure positivity of $k$ and $\varepsilon$.
In addition, if inflow is detected at any point on the far-field boundary, the boundary condition is switched from Neumann to Dirichlet by means of a suitably mollified indicator of the inflow region.
Finally, we stabilize the discrete equations using a strongly consistent stabilization technique (Galerkin Least Squares, GLS, stabilization) to address the convection-dominated nature of the RANS equations.
The complete formulation is shown in \cite{Villa17}.

The model equations described above are solved numerically using a finite element discretization.
The discretization is implemented in FEniCS~\cite{AlnaesBlechta2015a} by specifying the weak form of the residual, including the GLS stabilization and mollified versions of the positivity constraints on $k$ and $\epsilon$ and the switching boundary condition on the outflow boundary.
To solve the nonlinear system of equations that arise from the finite element discretization, we employ a damped Newton method.
The bilinear form of the state Jacobian operator is computed using FEniCS's symbolic differentiation capabilities.
Finally, we use pseudo-time continuation to guarantee global convergence of the Newton method to a physically stable solution (if such solution exists)~\cite{Kelley98}.
The finite element solver is detailed in \cite{Villa17}.

	\subsection{Uncertain parameters and quantity of interest} \label{sub:jet_uncertain}
	
	In this investigation five uncertain parameters are considered: velocity intensity at inlet\footnote{Since we keep other physical parameters constant, by varying the velocity intensity we effectively change the Reynolds number.} ($U$), mixing length at inlet ($\ell_m$), and the $k-\epsilon$ turbulence model coefficients $C_{\mu}$, $C_{2\epsilon}$, and $\sigma_k$:
	\begin{equation*}
		\param = [U, \ell_m, C_{\mu}, C_{2\epsilon}, \sigma_k].
	\end{equation*}
	The parameter domain is $\param \in \paramdom = [0.5, 1.5  ] \times [ 0.05, 0.15 ] \times [0.01, 0.15 ] \times [ 1.1, 2.5] \times [ 0.5, 2.5 ] $, and the nominal distribution of parameters is uniform in $\paramdom$.
	
	The other two coefficients of the $k-\epsilon$ turbulence model, $C_{1\epsilon}$ and $\sigma_{\epsilon}$, are also uncertain but do not vary independently.
	According to Dunn et al.~\cite{dunn:2011}, empirical evidence suggests that $C_{1\epsilon}$ is related to $C_{2\epsilon}$ by
	\begin{equation*}
		C_{1\epsilon} = \dfrac{C_{2\epsilon} - 0.8}{1.8}.
	\end{equation*}
	In addition, as noted in \cite{dunn:2011, guillas:2014}, the log-law implies that $\sigma_{\epsilon}$ must follow from
	\begin{equation*}
		\sigma_{\epsilon} = \dfrac{\kappa^2}{\sqrt{C_{\mu}}(C_{2\epsilon} - C_{1\epsilon})},
	\end{equation*}
	where $\kappa=0.41$ is the von K\'arman constant.
	
	\begin{figure}[h!]
		\begin{center}
			\subfigure[{$\param =[0.53, 0.07, 0.13, 1.20, 0.91]$}, $w=1.13$.]{\includegraphics[width=0.45\textwidth, trim={1.5in 0.5in 1.5in 0.5in}, clip]{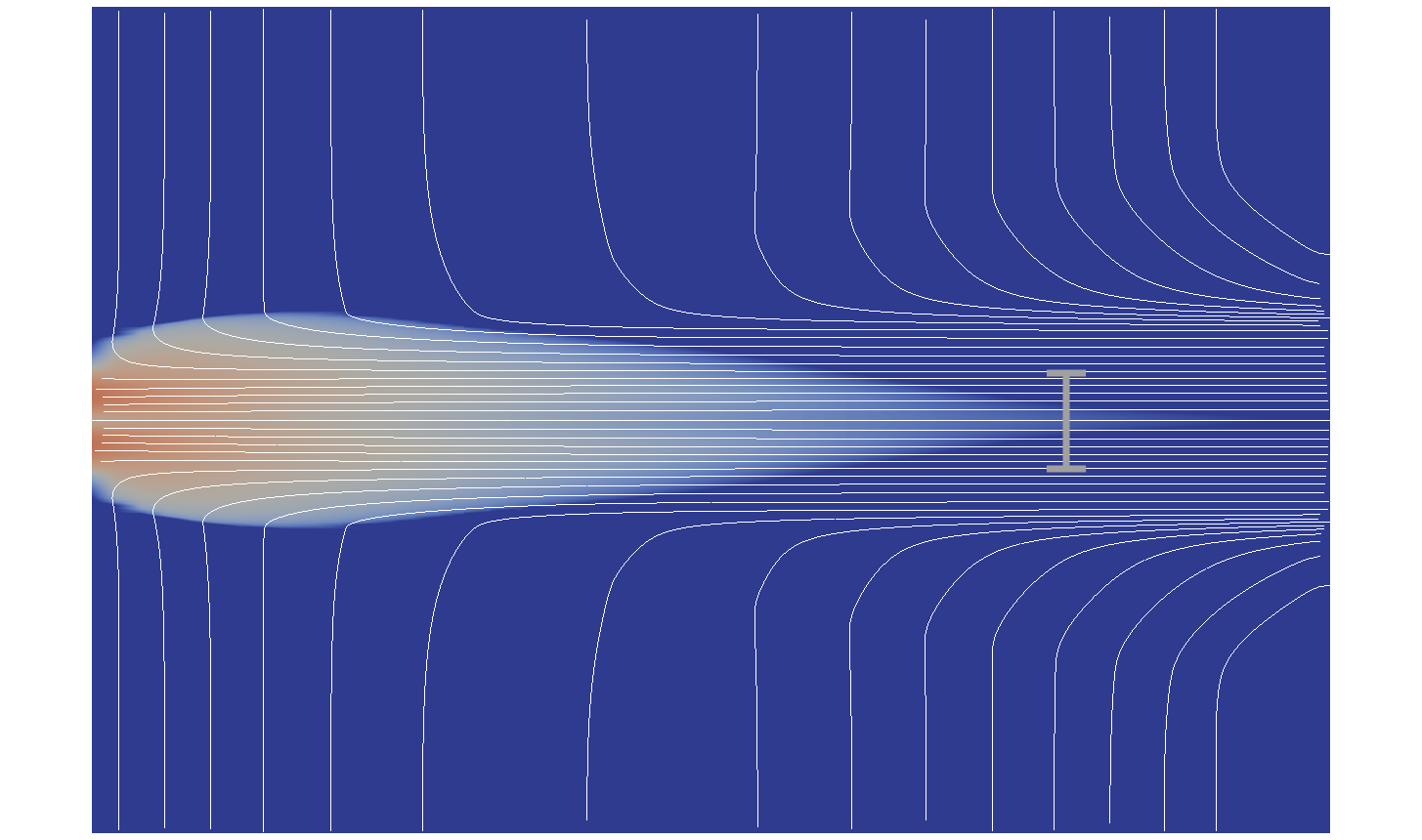}}
			\subfigure[{$\param =[0.86, 0.13, 0.08, 1.21, 1.08]$}, $w=2.38$.]
			{\includegraphics[width=0.45\textwidth, trim={1.5in 0.5in 1.5in 0.5in}, clip]{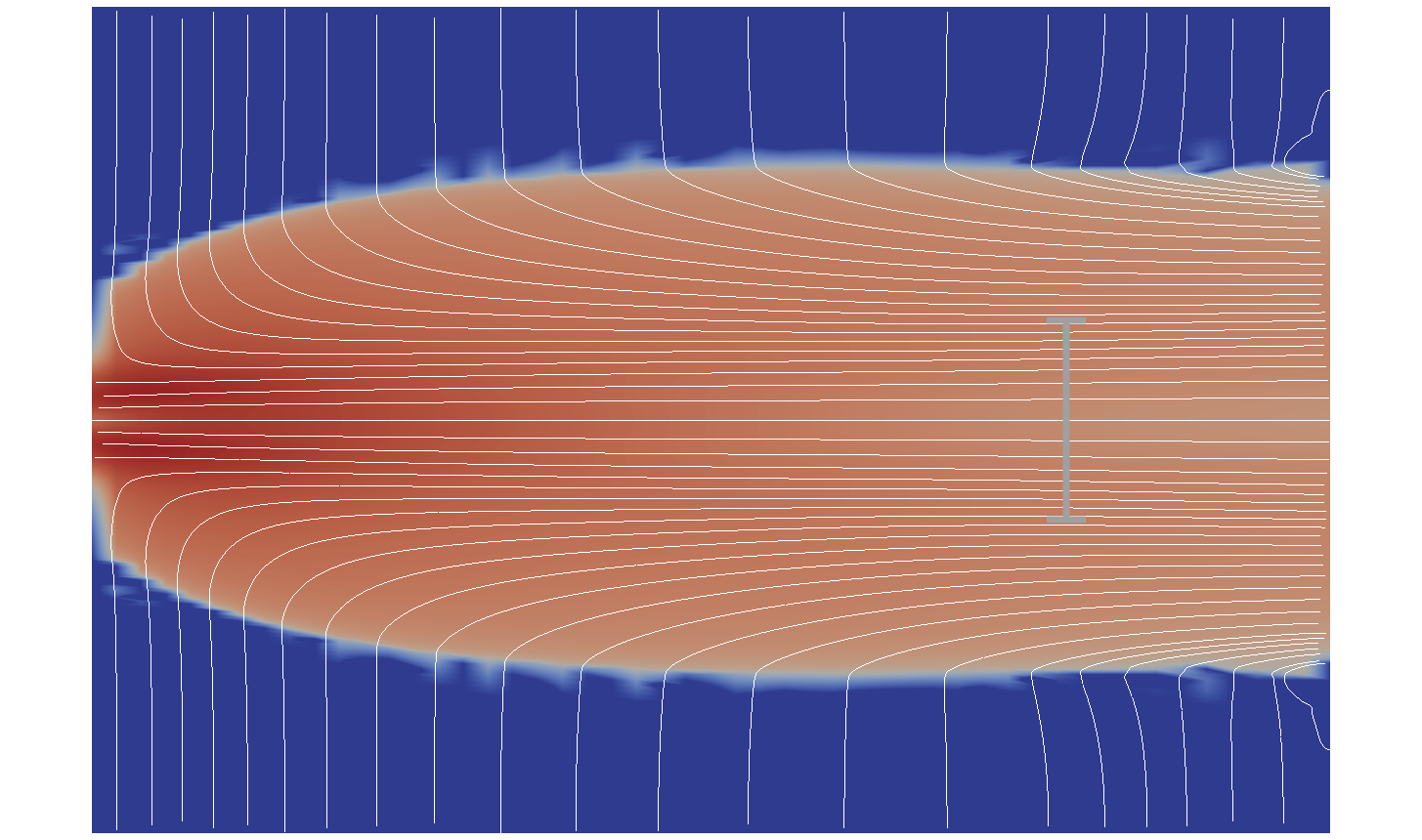}}\\
			\subfigure[{$\param =[0.96, 0.10, 0.08, 1.15, 1.59]$}, $w=2.78$.]{\includegraphics[width=0.45\textwidth, trim={1.5in 0.5in 1.5in 0.5in}, clip]{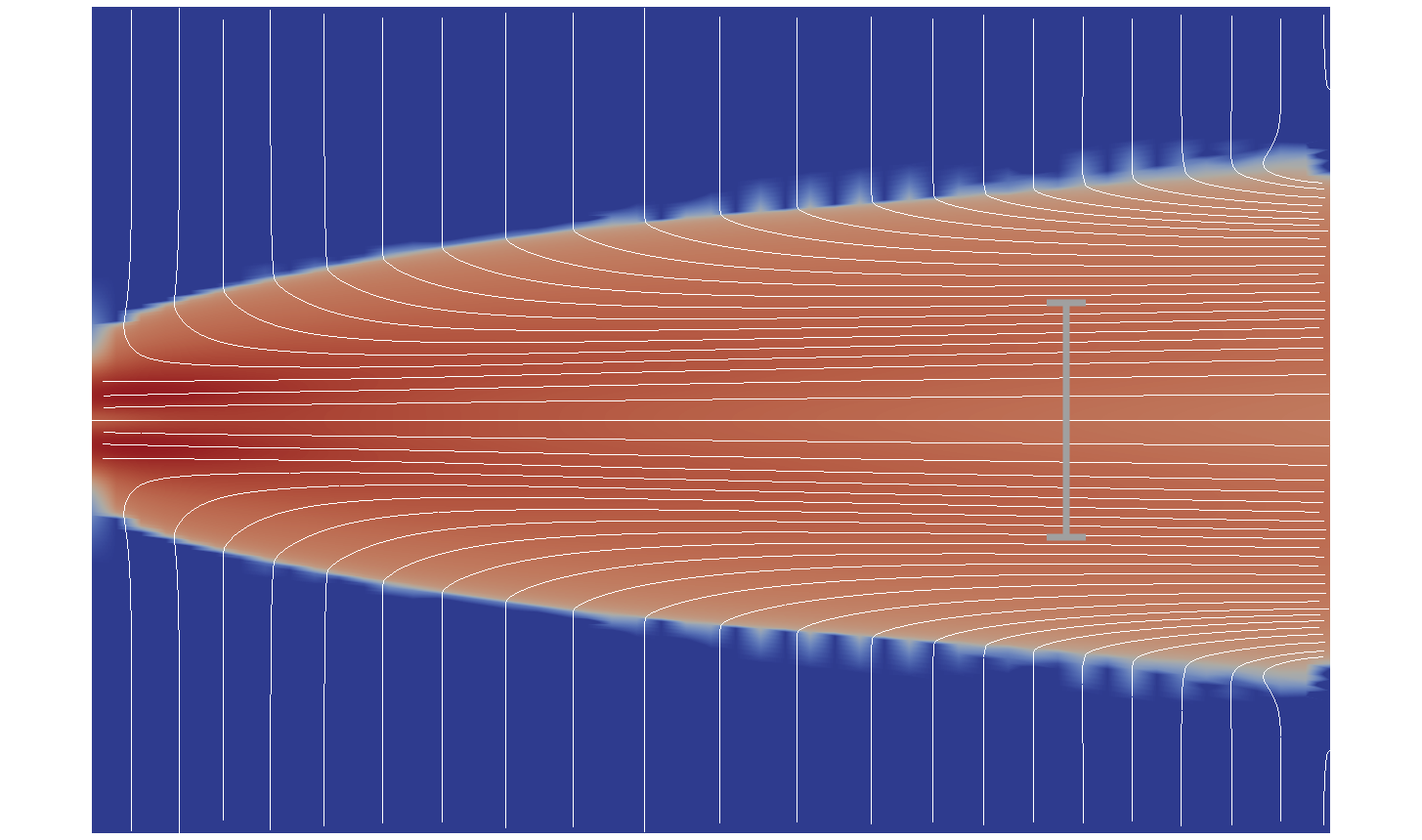}}
			\subfigure[{$\param =[1.11, 0.14, 0.12, 2.49, 0.69]$}, $w=4.13$.]{\includegraphics[width=0.45\textwidth, trim={1.5in 0.5in 1.5in 0.5in}, clip]{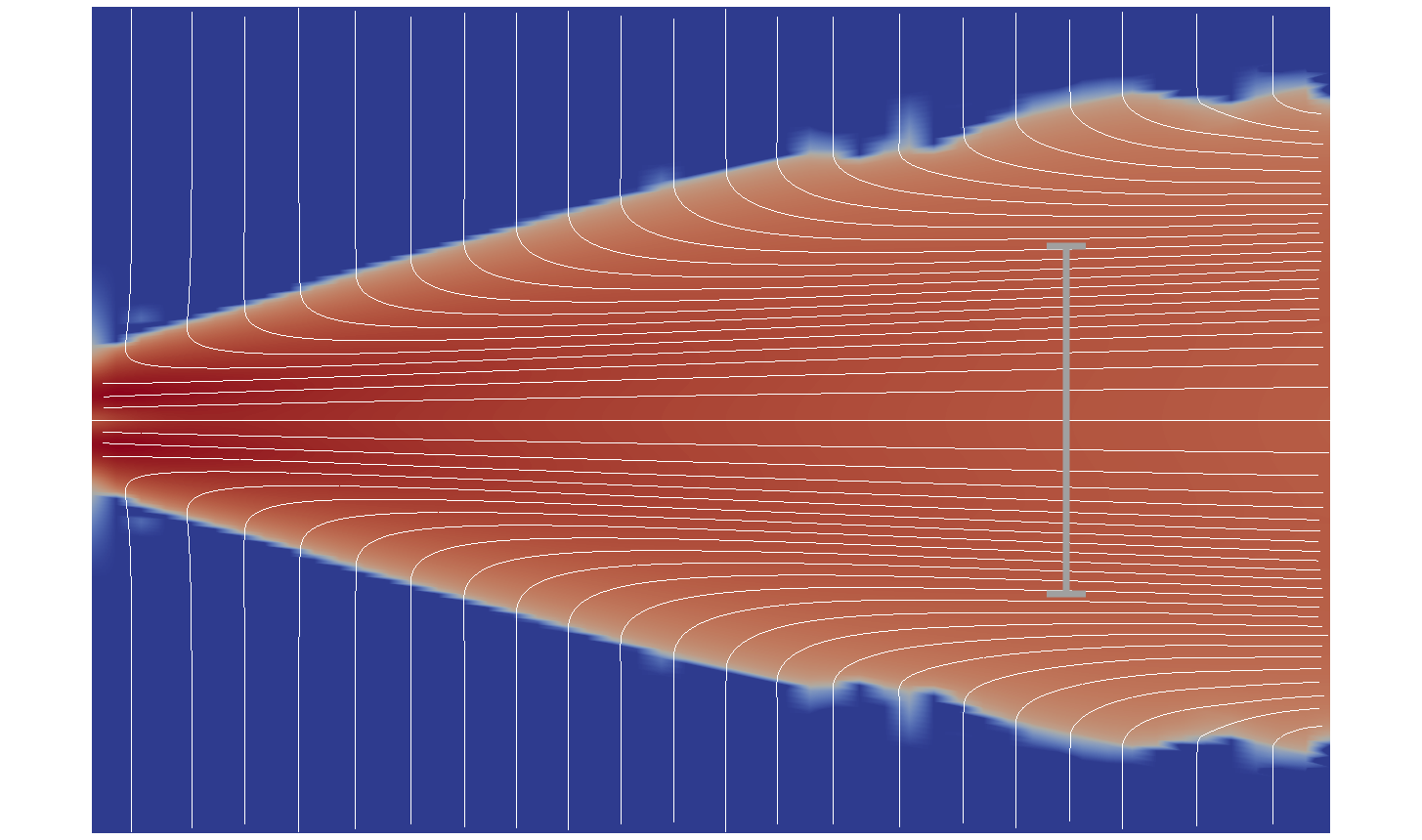}}
			\includegraphics[width=0.70\textwidth]{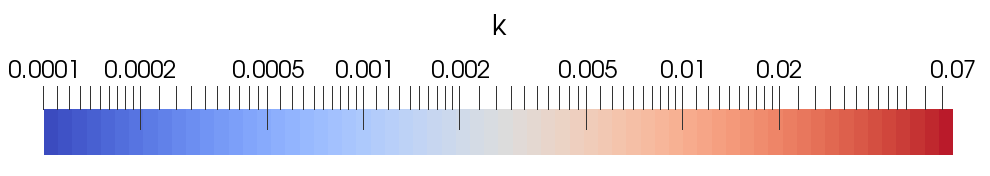}
		\end{center}
		\caption{Samples of the flow solution computed at different points of the input parameter space $\paramdom$.
				The plots show contours of turbulent kinetic energy and velocity streamlines. The white bars denote the integral jet width $w$ associated with each case.}
		\label{fig:jet_samples}
	\end{figure}
	
	The quantity of interest is the integral jet width measured at $x = 27.5D$:
	\begin{equation} \label{eq:w}
		w(\boldsymbol{v}; \param) = \dfrac{1}{v_{x_0}D}\int_0^{10D} v_x(x=27.5D,y; \param)\,dy,
	\end{equation}
	where $v_{x_0} = v_x(x=27.5D,y=0; \param)$.
	Figure~\ref{fig:jet_samples} illustrates a typical solution behavior for this turbulent jet by plotting contours of the turbulent kinetic energy for selected samples in $\paramdom$.

\subsection{Simplified-physics surrogate models} \label{sub:jet_surrogate}

We consider four surrogate models to represent the dynamics of the free plane jet flow.
The models are based on two distinct computational grids (fine and coarse), and on two representations of turbulence effects.
The fine computational grid contains 10,000 elements and 5,151 nodes, while the coarse grid contains 2,500 elements and 1,326 nodes. 
Furthermore, the models are based either on the complete $k-\epsilon$ turbulence model described in the previous section, or on a prescribed turbulent viscosity field.

In the latter case, the turbulent viscosity field is estimated by a linear interpolation based on 243 conditions that span the input parameter space $\mathcal{D}$ on a uniform grid (3 points along each of the 5 dimensions). At each of these 243 conditions, the turbulent viscosity field is computed with the $k-\epsilon$ turbulence model and the fine computational grid.

The following four low-fidelity models are increasingly complex in terms of either modeled physics or grid resolution:
\begin{itemize}
	\item \textbf{LFM1--CI}: Coarse, interpolated; combines the interpolated turbulence viscosity field with the coarse computational grid (3,978 degrees of freedom); average computational time $25$s;
	\item \textbf{LFM2--FI}: Fine, interpolated; combines the interpolated turbulence viscosity field with the fine computational grid (15,453 degrees of freedom); average computational time $72$s;
	\item \textbf{LFM3--CKE}: Coarse $k-\epsilon$; combines the $k-\epsilon$ turbulence model with the coarse computational grid (6,630 degrees of freedom); average computational time $109$s;
	\item \textbf{HFM}: High-fidelity model; combines the $k-\epsilon$ turbulence model with the fine computational grid (25,755 degrees of freedom); average computational time $590$s.
\end{itemize}

Note that the models based on an interpolated turbulent viscosity field run four to eight times faster than the corresponding models based on the $k-\epsilon$ turbulence model.

This speedup results from eliminating \eqref{eq:jet_k}--\eqref{eq:jet_e} from the governing equations, which leads to a reduction in the total number of degrees of freedom (elimination of variables $k$ and $\epsilon$) and simplifications in the numerical discretization.

Let $\boldsymbol{v}_i$, $i = $ HFM, LFM1, LFM2, LFM3, denote the velocity field computed with the models above.
The high-fidelity model is the mapping from the inputs to the quantity of interest (jet width from \eqref{eq:w}) for a velocity field computed with the most complex representation of the flow dynamics, $\boldsymbol{v}_{\text{HFM}}$:
\begin{equation*}
	f: \mathcal{D} \mapsto \mathbb{R}, \qquad f(\param) = w(\boldsymbol{v}_{\text{HFM}}; \param).
\end{equation*}
The surrogate models are defined in a similar fashion as
\begin{equation*}
	f^{(i)}: \mathcal{D} \mapsto \mathbb{R}, \qquad f^{(i)}(\param) = w(\boldsymbol{v}_i; \param), \quad i = \text{LFM1, LFM2, LFM3}.
\end{equation*}

	\subsection{Results for multifidelity fusion of small probability estimators}
	\label{sub:jet_results}

	We define a design failure when the jet width is below the value 0.98.
	Hence, the limit state function is given by
	\begin{equation}
		g(f(\mathbf{z})) = f(\mathbf{z}) - 0.98. 
	\end{equation}
	We compute the biasing distributions $q_{i}$ for $ i=\text{LFM1, LFM2, LFM3}$ from the three low-fidelity surrogate models  via MFIS (see Section~\ref{sec:MFIS}). For each surrogate, we draw $\hat{m}=20,000$ parameter samples from the uniform distribution on $\mathcal{D}$ and evaluate the limit state function applied to the resulting quantity of interest. If the limit state function indicates failure of the system for a solution obtained from the $i$th surrogate model, the corresponding parameter is added to $\mathcal{G}^{(i)}$, the failure set computed from the $i$th surrogate model. We then fit a multivariate Gaussian to the samples in $\mathcal{G}^{(i)}$, resulting in the biasing densities $q_\text{LFM1}, q_\text{LFM2}, q_\text{LFM3}$. 
	
	Evaluation of the limit state function with the threshold value of 0.98 resulted in few samples in the failure region, so we increased it to 1.12 to obtain more samples to compute the biasing density from. For the three surrogate models and the high-fidelity model, the $\hat{m}=20,000$ evaluations yield 21, 21, 30 and 76 samples, respectively, where the QoI falls below that increased threshold. This strategy yields an efficient biasing density as we see below.  
	As reference, we repeat the same process with the high-fidelity model, resulting in the biasing distribution $q_{\text{HFM}}$. 
	
\begin{figure}[H]
		\begin{center}
			\includegraphics[width=0.49\textwidth]{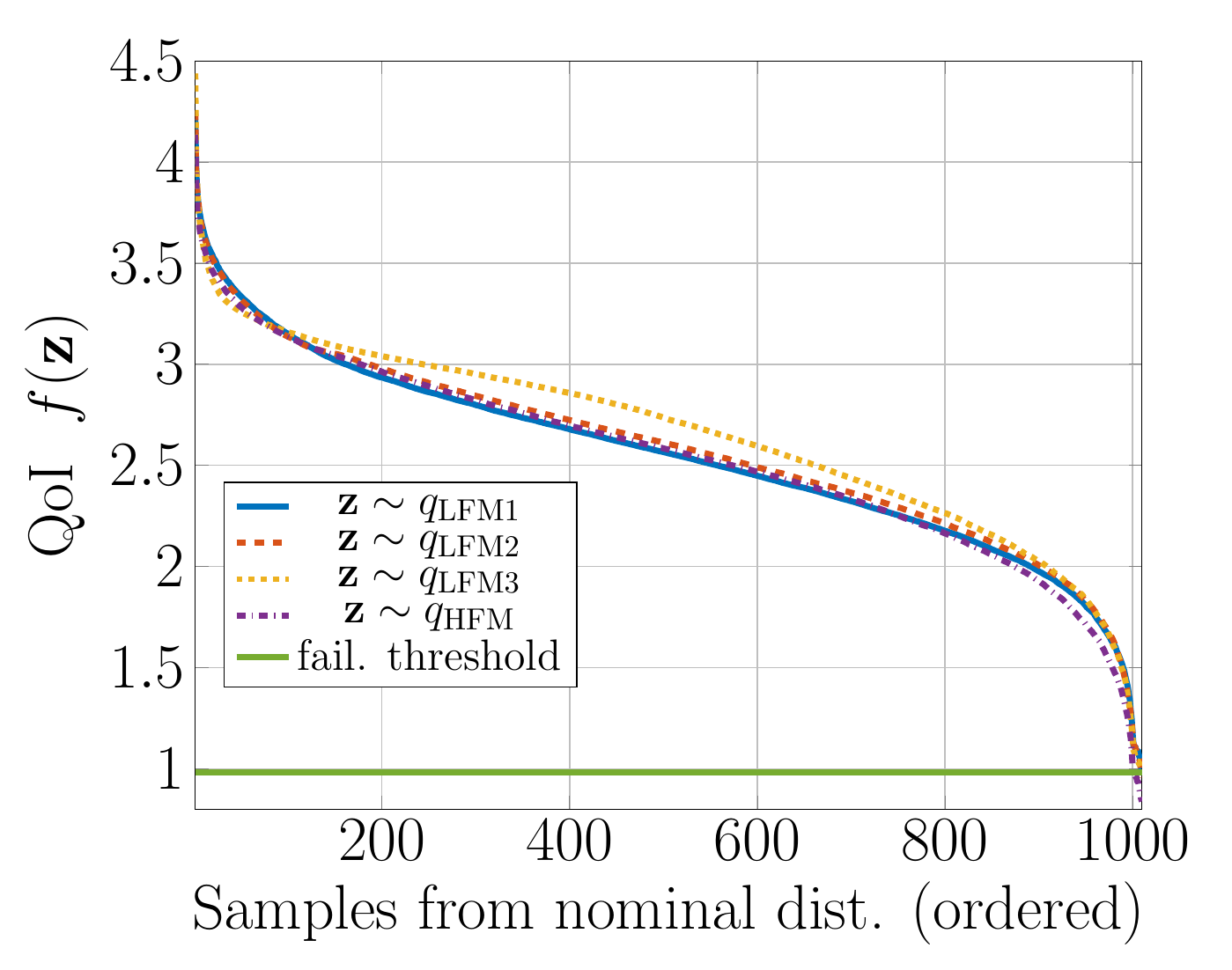}
			\includegraphics[width=0.49\textwidth]{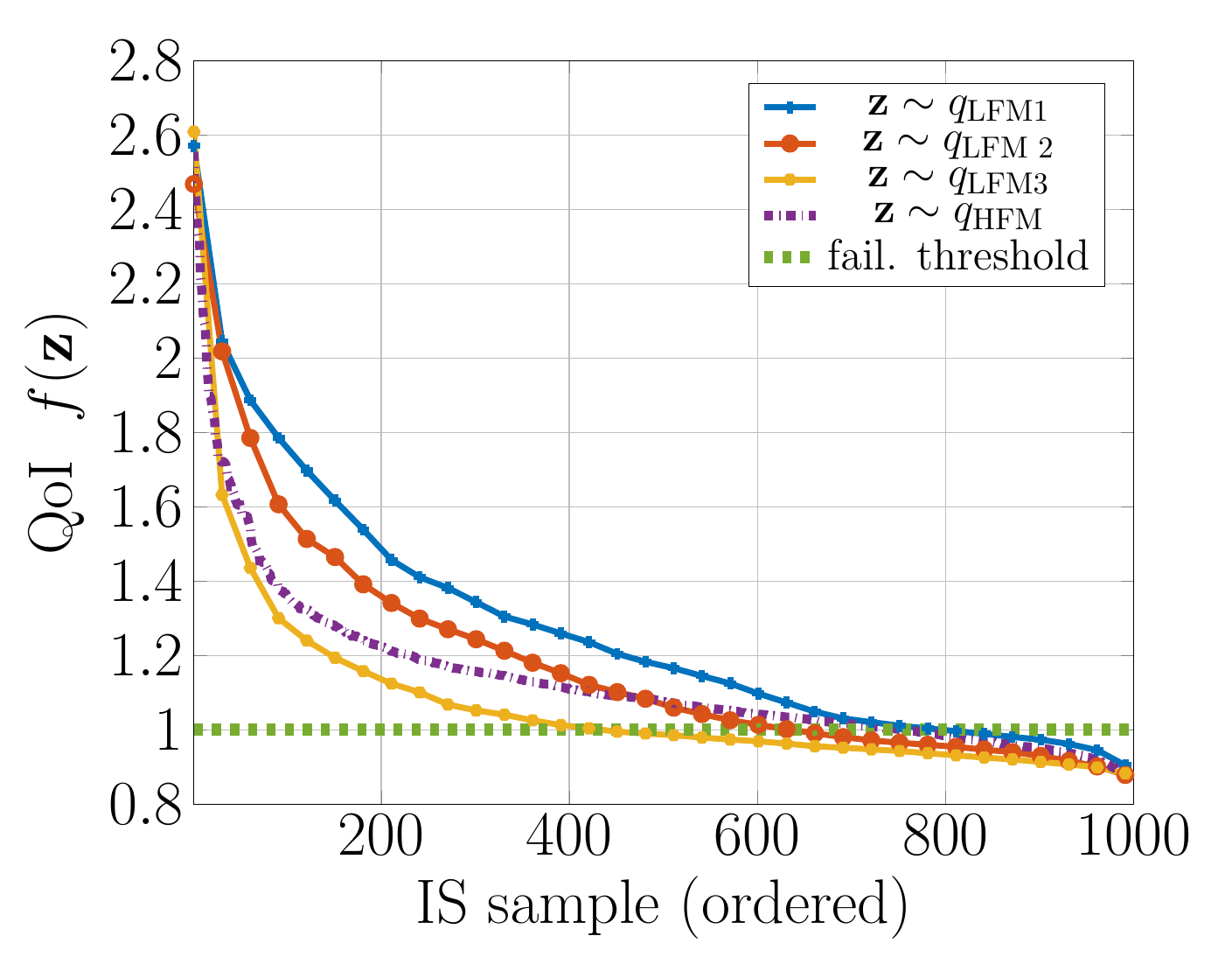}
			\caption{Quantity of interest, the width of the jet at $x=27.5D$, for $n=10^3$ samples. Left: The input parameters are drawn from the nominal distribution. Right: The input parameters are drawn from different biasing distributions. Note the large portion of samples falling below the failure threshold, and the different scaling of the $y$-axis.}
			\label{fig:jet_QoI}
		\end{center}
	\end{figure}

	First, we investigate the quality of the biasing distributions. For reference, Figure~\ref{fig:jet_QoI}, left, shows the result of $10^3$ uniform sample evaluations with the four computational models. Note that hardly any samples are below the failure threshold. In contrast, the quantity of interest computed from samples of the four biasing distributions is shown in Figure~\ref{fig:jet_QoI}, right. The biasing distributions give between 10\%-50\% of the 1000 drawn samples in the failure domain. 
	Note, that the y-axis scaling of both figures is different, which also shows that the biased samples result in a tighter range of QoI values than the unbiased samples. Thus, the biasing distributions are indeed biased towards the failure region, and therefore the multifidelity strategy provides a viable way of saving computational time to inform a biasing distribution.

	The reference failure probability is computed via importance sampling with ${n=10^4}$ samples drawn from the HFM biasing distribution  and is ${\hat{P}_{7,500,q_{\text{HFM}}}^{\text{IS}} =  7.25\times 10^{-4}}$. 
	We compute the estimators $P_{n_i}^{\text{IS}}, i=1,2,3$ with $n_i$ samples using the biasing densities $q_\text{LFM1}, q_\text{LFM2}, q_\text{LFM3}$ derived from the three surrogate models. We obtain the fused multifidelity estimator $\Palpha$ as described above in Algorithm~\ref{alg:MIS} with $n_i=\lfloor n/3 \rfloor, i=1,2,3,$ samples by fusing the three surrogate-model-based importance sampling estimators. The fused estimator thus uses a total of $n$ samples. We compare these estimators with an estimator that uses $n$ samples from the HFM biasing density $q_{\text{HFM}}$. The estimators and the error measures are averaged over three independent runs. 

\begin{figure}[H]
		\begin{centering}
			\includegraphics[width=0.49\textwidth]{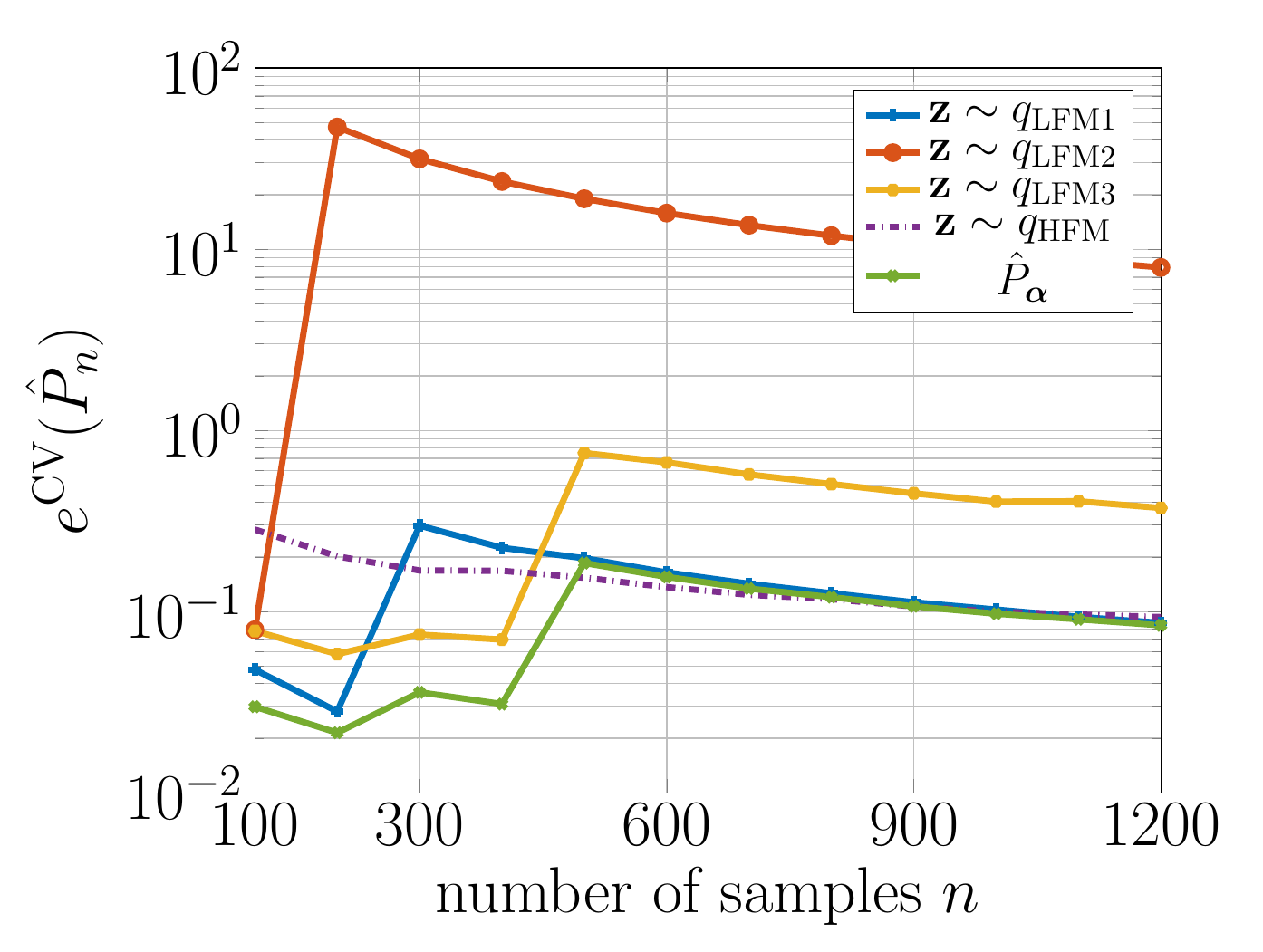}
			\includegraphics[width=0.49\textwidth]{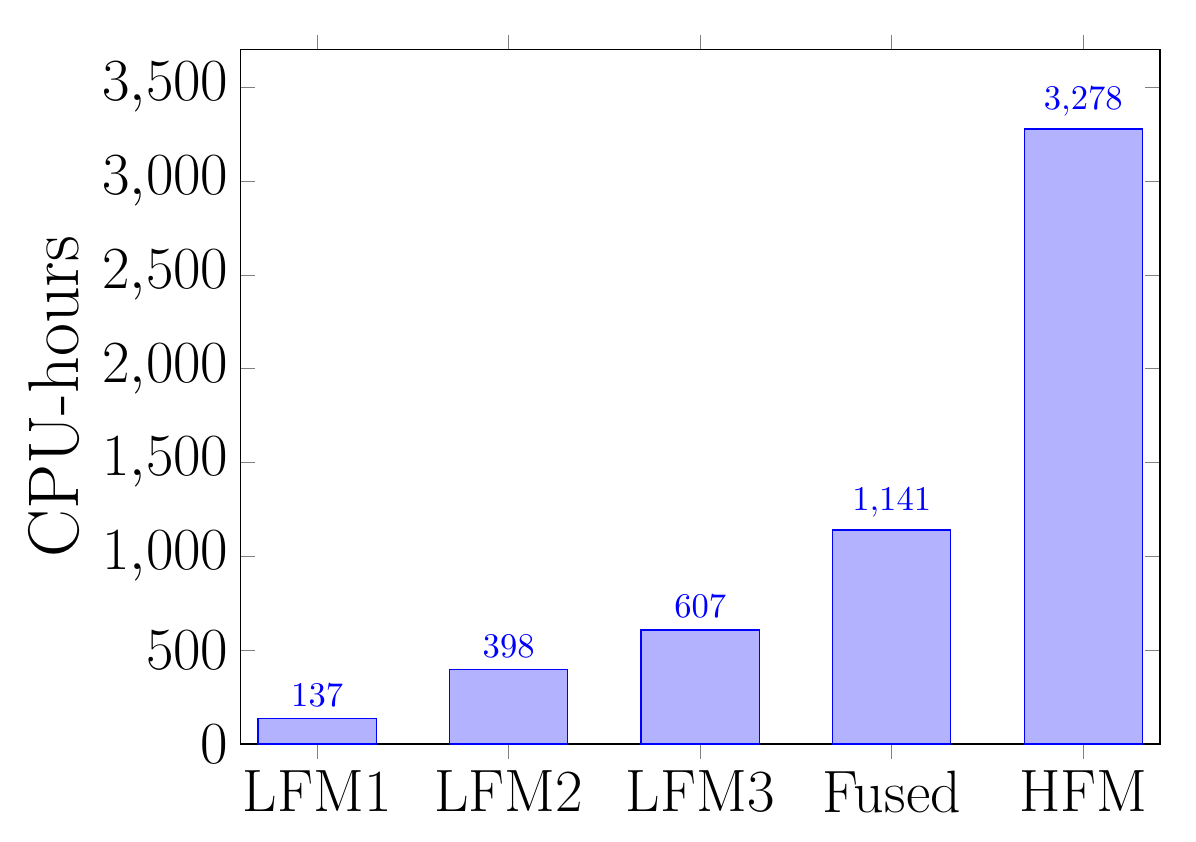}
			\caption{Free plane jet application with five uncertain parameters. Left: Coefficient of variation for the different estimators. Right: CPU-hours for construction of the biasing densities used in the estimators in the left plot. }
			\label{fig:jet_RSME}
		\end{centering}
	\end{figure}

	The coefficient of variation \eqref{eq:coefvar} is shown in Figure~\ref{fig:jet_RSME}, left. The biasing density derived from the high-fidelity model yields the best estimator among all the models, as expected. The fused estimator yields a better coefficient of variation than LFM2 and LFM3, shows almost identical convergence as the estimator using $q_\text{LFM1}$. 
	Table~\ref{tbl:alpha} shows the three weights for the fused estimator $\hat{P}_{\boldsymbol{\alpha}}$ as given in Proposition~\ref{prop:weights}, according to which, the estimates with the lowest variance get assigned the largest weights. 
	
	\begin{table}[!ht]
		\centering
		\caption{Weights of the fused estimator $\hat{P}_{\boldsymbol{\alpha}}$ with $n$ samples.}
		\begin{tabular}{l | c c c c c}
			&  $n=300$ &  $n=600$& $n=900$& $n=1200$ \\
			\hline
			$\alpha_1$  &   0.500 &   0.502  &  0.610  &  0.900\\
			$\alpha_2$  &   0.448 &   0.044  &  0.055  &  0.057\\
			$\alpha_3$  &   0.052 &   0.453  &  0.335  &  0.043\\
		\end{tabular}
		\label{tbl:alpha}
	\end{table}
	
	The CPU-hours to compute the biasing densities via this approach are shown in Figure~\ref{fig:jet_RSME}, right. Since MC methods are embarrassingly parallel, any practical implementation can take advantage of this. Our numerical experiments were parallelized on a computing cluster with 55 nodes. Each node is a quad-core Intel Xenon E5-1620 with 3.6 GHz and 10MB Cache. The nodes have either 32GB or 64GB RAM.
	To put the CPU-hours savings achieved by using the high-fidelity model versus the lower-fidelity models to construct biasing densities into perspective, we see that using LFM1 reduces the computational cost by 96\%, LFM2 by 88\% and LFM3 by 81.5\%. If we are using a fused estimator of all three models, we still save more than 65\% computational effort compared to using the HFM, see Figure~\ref{fig:jet_RSME}, right. 
	This significant time difference can have important implications for engineering practice, as it translates into faster evaluation time and savings in CPU-hours . 
	
\section{Conclusions} \label{sec:conclusions}
We enabled the estimation of small probabilities for expensive-to-evaluate models via a new approach drawing from importance sampling, multifidelity modeling and information fusion. The effectiveness of our proposed approach is demonstrated on a convection-diffusion-reaction PDE, where asymptotic numerical results could be obtained. The strength of the proposed framework is then shown on the target application of the turbulent jet, a challenging problem for small-probability computation due to its high computational cost. 
The proposed framework was illustrated for the special case of importance-sampling based estimators, but applies to a much broader class of estimators, as long as the estimators are unbiased. An investigation of correlated estimators and the effect of correlation for variance reduction would be an interesting future direction. 
By fusing different estimators, we avoid the difficult biasing density selection problem. We also showed that this strategy always outperforms sampling from the worst biasing density. The numerical results suggest that the fused estimator is often comparable to an estimator that samples from the best biasing density only.

\section*{Acknowledgements}
The authors thank Prof. M. Klein for sharing the DNS data in \cite{klein15} with us.
This work was supported by the Defense Advanced Research Projects Agency [EQUiPS program, award  W911NF-15-2-0121, Program Manager F. Fahroo]; the Air Force [Center of Excellence on Multi-Fidelity Modeling of Rocket Combustor Dynamics, award FA9550-17-1-0195]; and the US Department of Energy, Office of Advanced Scientific Computing Research (ASCR) [Applied Mathematics Program, awards DE-FG02-08ER2585 and DE-SC0009297, as part of the DiaMonD Multifaceted Mathematics Integrated Capability Center].

\bibliographystyle{abbrv}
\bibliography{fusedIS.bib}

\end{document}